\documentclass{aa}  

\usepackage{showyourwork}

\usepackage{graphicx}
\usepackage{txfonts}
\usepackage[utf8]{inputenc} 
\usepackage[T1]{fontenc}    
\usepackage{siunitx}
\usepackage{hyperref}
\usepackage{url}
\usepackage{amsfonts}       
\usepackage{nicefrac}       
\usepackage{microtype}      
\usepackage{amsmath}
\usepackage{amssymb}
\usepackage{booktabs}       
\usepackage{tabularx}
\usepackage{tikz}

\usepackage{xcolor}

\usepackage{subfigure}

\begin{document}

   \title{Inferring Galactic Parameters from Chemical Abundances with Simulation-Based Inference}
   
   \titlerunning{Simulation-Based Inference for Galactic Parameters}
   \authorrunning{T. Buck and B. Günes}


   \author{Tobias Buck\inst{1,2}
          \and
          Berkay Günes\inst{1,2}
          \and
          Giuseppe Viterbo\inst{1,2}
          \and
          William H. Oliver\inst{1,2}
          \and
          Sven Buder\inst{3,4}
          }

   \institute{Interdisciplinary Center for Scientific Computing (IWR), University of Heidelberg,
 Im Neuenheimer Feld 205, D-69120 Heidelberg, Germany
 \and
 Universität Heidelberg, Zentrum für Astronomie, Institut für Theoretische Astrophysik, Albert-Ueberle-Straße 2, D-69120 Heidelberg, Germany
 \and
 Research School of Astronomy and Astrophysics, Australian National University, Canberra, ACT 2611, Australia
 \and
 ARC Centre of Excellence for All Sky Astrophysics in 3 Dimensions (ASTRO 3D), Australia\\
 \email{tobias.buck@iwr.uni-heidelberg.de}
             }

   \date{Received Month, XXXX; accepted Month Day, XXXX}

 \abstract
   {Galactic chemical abundances provide crucial insights into fundamental galactic parameters, such as the high-mass slope of the initial mass function (IMF) and the normalization of Type Ia supernova (SN\,Ia) rates. Constraining these parameters is essential for advancing our understanding of stellar feedback, metal enrichment, and galaxy formation processes. However, traditional Bayesian inference techniques, such as Hamiltonian Monte Carlo (HMC), are computationally prohibitive when applied to large datasets of modern stellar surveys.}
   {We leverage simulation-based-inference (SBI) as a scalable, robust, and efficient method for constraining galactic parameters from stellar chemical abundances and demonstrate its the advantages over HMC in terms of speed, scalability, and robustness against model misspecifications.}
   {We combine a Galactic Chemical Evolution (GCE) model, \texttt{CHEMPY}, with a neural network emulator and a Neural Posterior Estimator (NPE) to train our SBI pipeline. Mock datasets are generated using \texttt{CHEMPY}, including scenarios with mismatched nucleosynthetic yields, with additional tests conducted on data from a simulated Milky Way-like galaxy. SBI results are benchmarked against HMC-based inference, focusing on computational performance, accuracy, and resilience to systematic discrepancies.}
   {SBI achieves a $\sim75,600\times$ speed-up compared to HMC, reducing inference runtime from $\gtrsim42$ hours to mere seconds for thousands of stars. Inference on $1,000$ stars yields precise estimates for the IMF slope ($\alpha_{\rm IMF} = -2.298 \pm 0.002$) and SN\,Ia normalization ($\log_{10}(N_{\rm Ia}) = -2.885 \pm 0.003$), deviating less than 0.05\% from the ground truth. SBI also demonstrates similar robustness to model misspecification than HMC, recovering accurate parameters even with alternate yield tables or data from a cosmological simulation.}
   {SBI represents a paradigm shift in GCE studies, enabling efficient and precise analysis of massive stellar datasets. By outperforming HMC in speed, scalability, and robustness, SBI is poised to become a cornerstone methodology for future spectroscopic surveys facilitating deeper insights into the chemical and dynamical evolution of galaxies.}


\maketitle
\section{Introduction}

Understanding the chemical enrichment of galaxies is fundamental to deciphering their formation and evolution. Chemical abundances of stars offer a wealth of information about galactic parameters, such as the high-mass slope of the initial mass function (IMF) and the normalization of Type Ia supernova (SN\,Ia) rates. These parameters critically influence the production of heavy elements \citep[e.g.][]{2005A&A...430..491R,2015MNRAS.449.1327V,2015MNRAS.451.3693M}, stellar feedback, and star formation histories, making their accurate determination essential for realistic hydrodynamical simulations of galaxy formation \citep[e.g.][]{Sawala2016,Hopkins2018,Pillepich2018,Buck2020,Buck2020c,Buck2021,Font2020,Agertz2021}. Despite their importance, constraining these parameters has proven challenging due to limited observational data and the computational demands of traditional inference techniques.

For example, a range of high-mass IMF slopes have been suggested \citep[Tab.\,7]{2016ApJ...824...82C}, with a steeper-than-canonical slope being suggested by a range of studies \citep[e.g.][]{2015ApJ...806..198W,Rybizki2015,Chabrier2014}. In addition, the IMF slope may itself be not a constant but rather a function of metallicity, introducing further complexity \citep[e.g.][]{2019MNRAS.482..118G,Martin2019}. Similarly, the choice of SN\,Ia delay-time-distribution and normalization plays a crucial role in the enrichment of the interstellar medium \citep[ISM; e.g.][]{Buck2021} and is heavily debated \citep{2010ApJ...722.1879M,2012MNRAS.426.3282M,2015ApJ...810..137J}.

\begin{figure*}[]
     \centering
     \includegraphics[width=1\linewidth]{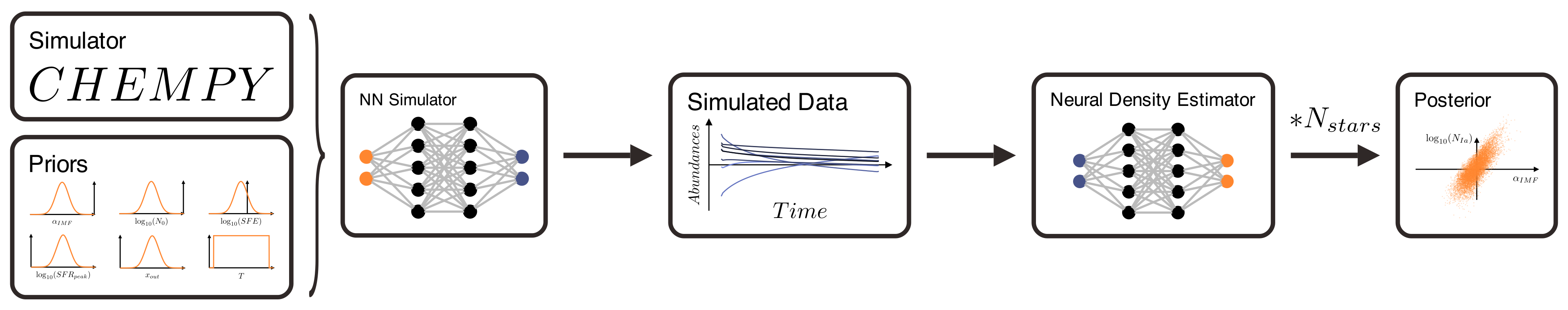}
     \vspace{-.5cm}
     \caption{SBI flow chart. From a set of priors we simulate a sample of stellar abundances using \texttt{CHEMPY} \citep{Rybizki_2017,Philcox_2019} which we use to train a \emph{neural network} emulator to speed up the data generation process. Using the \emph{neural network} emulator we produce training data to train the Neural Density Estimator. With this we infer the posterior distribution of the model parameters from a single star. Repeating that for $N_{\rm stars}$ from the same galaxy gives an accurate fit of the IMF slope and Type Ia supernovae normalization.}
     \label{fig:flowchart}
\end{figure*}

Recent advances in stellar spectroscopic surveys, such as APOGEE \citep{apogee17} and GALAH \citep{Buder2021,Buder2024}, have produced unprecedented datasets of stellar chemical abundances across a third of the period table. These datasets hold the potential to unlock detailed constraints on galactic parameters across diverse environments. However, traditional Bayesian inference methods, such as Markov Chain Monte Carlo (MCMC) and Hamiltonian Monte Carlo (HMC), struggle to scale to these large datasets. Such methods are computationally expensive, requiring hours of runtime for even modest sample sizes, and are susceptible to biases when confronted with high-dimensional posterior distributions.

In this work, we present a novel approach leveraging simulation-based inference \citep[SBI, e.g.][]{Cranmer2020} to address these limitations. SBI bypasses the need for explicit likelihood functions, enabling efficient and scalable inference of galactic parameters directly from simulated stellar abundances. By combining a neural network emulator for the \texttt{CHEMPY} Galactic Chemical Evolution (GCE) model with a Neural Posterior Estimator (NPE), we achieve rapid and robust inference. Unlike HMC, which requires extensive sampling for each dataset, our method amortizes the computational cost during training, allowing subsequent inference to scale seamlessly to larger datasets.

This study focuses on two critical global galactic parameters: 
the high-mass slope of the \citet[Tab.\,1]{2003PASP..115..763C} IMF ($\alpha_{\rm IMF}$) and the SN\,Ia normalization, $\log_{10}(N_{\rm Ia})$, the rate of SN\,Ia explosions per unit mass. We demonstrate the accuracy, scalability, and robustness of SBI through tests on mock datasets generated by \texttt{CHEMPY} \citep{Rybizki_2017}, as well as on data from hydrodynamical simulations. Additionally, we compare our results to those obtained using HMC-based inference on the same datasets \citep[see][]{Philcox_2019}, highlighting SBI's superior performance in terms of speed, precision, and resilience to model misspecification.

The structure of this paper is as follows: In Section~\ref{sec:methods}, we outline the methods used, including the GCE model and SBI framework. Section~\ref{sec: Results} presents our results on both \texttt{CHEMPY} and IllustrisTNG \citep{Pillepich2018} mock data, emphasizing SBI's advantages over traditional approaches. Finally, in Section~\ref{sec: discussion}, we discuss the broader implications of our findings and outline potential future applications, before concluding in Section~\ref{sec: conclusion}.

Finally, we publicly release all of our code to reproduce the results of this manuscript via GitHub\footnote{URL: {\url{https://github.com/TobiBu/sbi-chempy}}} and refer to Appendix \ref{sec:appendix_code_and_data} for a more extended overview of our code availability. All our datasets and network weights are publicly available on Zenodo.\footnote{URL: \url{https://zenodo.org/records/14925307}}

\section{Methods}
\label{sec:methods}

In order to establish our new method based on SBI we need two ingredients: A simulator (in our case a GCE model) that simulates observational data from a set of model parameters (in our case the IMF slope and the Type Ia supernovae normalization) and a flexible way of parametrizing the posterior density conditioned on the observation in order to perform our inference (see Fig. \ref{fig:flowchart} for a schematic visual representation of our method). In the next subsections we describe both ingredients in detail.

\subsection{Galactic chemical evolution models}
Our simulator is based on the \texttt{CHEMPY} model \citep{Rybizki_2017}. \texttt{CHEMPY} is a simple GCE model that is able to predict stellar chemical abundances throughout cosmic time by using published nucleosynthetic yield tables for three key processes (SN\,Ia and SN\,II explosions and AGB stellar feedback) and a small number of parameters controlling simple stellar populations (SSPs) and ISM physics. We refer to the initial \texttt{CHEMPY} paper \citep{Rybizki_2017} for the details of the model.

\begin{tiny}
\begin{table*}
\begin{minipage}{\textwidth}
\begin{center}
\caption{Free \texttt{CHEMPY} parameters for each star, with their prior values and Gaussian widths. Stellar birth-times are set for each star individually from a Uniform prior, based on realistic age estimates.}
\begin{tabularx}{\textwidth}{ >{\raggedleft}p{2.2cm}p{6.5cm}|c c }
Parameter & Description & $\overline{\theta}_\mathrm{prior}\pm\sigma_\mathrm{prior}$ & Prior from: \\

\hline
\multicolumn{4}{c}{$\vec{\Lambda}$: \textit{Global stellar (SSP) parameters}}\\
\hline
$\alpha_\mathrm{IMF}$ & High-mass slope of the \citep{2003PASP..115..763C} IMF & $-2.3\pm0.3$ & \citep[Tab.\,1]{2003PASP..115..763C} \\
  
$\log_{10}\left(N_\mathrm{Ia}\right)$ & Number of SN\,Ia per $\mathrm{M}_\odot$ over 15\,Gyr & $-2.89\pm0.3$ & \citep[Tab.1\,]{2012PASA...29..447M}\\
  
\hline
\multicolumn{4}{c}{$\vec{\Theta}_i$: \textit{Local ISM parameters}}\\
  
\hline
$\log_{10}\left(\mathrm{SFE}\right)$ & Star formation efficiency governing gas infall & $-0.3\pm0.3$ & \citep{2008AJ....136.2846B}\\
  
$\log_{10}\left(\mathrm{SFR}_\mathrm{peak}\right)$ & SFR peak in Gyr (scale of $k=2$ $\Gamma$-distribution) & $0.55\pm0.1$ & \citep[fig.\,4b]{2013ApJ...771L..35V}\\
  
x$_\mathrm{out}$ & Stellar feedback fraction & $\phantom{-}0.5\pm0.1$ & \citep[Tab.\,1]{Rybizki_2017}\\
  
\hline
\multicolumn{4}{c}{$T_i$: \textit{Timescale}}\\
 
\hline
$T_i$ & Time of stellar birth in Gyr & [$1$,$13.8$] & Observations

\label{tab:priors}
\end{tabularx}
\end{center}
\end{minipage}
\end{table*}
\end{tiny}

In particular, we are using the \texttt{CHEMPYScoring} module \citep{Philcox_2018} publicly available as the \texttt{CHEMPYMulti} \citep{Philcox_2019}\footnote{\href{https://github.com/oliverphilcox/ChempyMulti}{github.com/oliverphilcox/ChempyMulti}} package a further development of the original \texttt{CHEMPY} model. 

\paragraph{\texttt{CHEMPY} parameters:}
In this work, we allow six \texttt{CHEMPY} parameters to vary freely (see also Tab.\,\ref{tab:priors}). These can be categorized into three groups:
\begin{enumerate}
     \item $\vec\Lambda$: \textbf{Global Galactic Parameters} describe SSP physics and comprise the high-mass \citet{2003PASP..115..763C} IMF slope, $\alpha_\mathrm{IMF}$, which effectively sets the number of CC-SNe a SSP generates and (logarithmic) Type Ia SN normalization, $\log_{10}(N_\mathrm{Ia})$, which controls the total number of SN Is per SSP. We treat these as star-independent and assume them to be constant across galactic environments and cosmic time\footnote{Whilst $\log_{10}(N_\mathrm{Ia})$ is constant with respect to time by definition, it being simply a normalization constant, there is some evidence for $\alpha_\mathrm{IMF}$ varying as a function of time or metallicity \citep{Chabrier2014,2016MNRAS.462.2832C,2019MNRAS.482..118G,Martin2019}.}. 
     We adopt the same broad priors as \citep{Philcox_2019} for these variables (see also Tab.\ref{tab:priors}). 
     \item $\{\vec\Theta_i\}$: \textbf{Local Galactic Parameters} describe the local physics of the ISM and are hence specific to each stellar environment, indexed by $i$. As defined in \citep{Rybizki_2017}, these include the star-formation efficiency (SFE), $\log_{10}(\text{SFE})$, which qunatifies the star formation rate per unit gas, the peak of the star formation rate (SFR), $\log_{10}(\mathrm{SFR}_\mathrm{peak})$, and the fraction of stellar outflow that is fed to the gas reservoir, $\mathrm{x}_\mathrm{out}$. We adopt broad priors for all parameters and, as in \citep{Philcox_2019}, fix the SN\,Ia delay-time distribution, $\log_{10}(\tau_{\rm Ia})$, to $\log_{10}(\tau_{\rm Ia})=-0.80$ \citep[see also][]{Philcox_2018}.
     \item $\{T_i\}$: \textbf{Stellar Birth-Times}. Time in Gyr at which a given star is formed from the ISM. We assume that its proto-stellar abundances match the local ISM abundances at $T_i$.
\end{enumerate}

The separability of local (ISM) parameters and global (SSP) parameters is motivated by recent observational evidence: \citet{2019arXiv190710606N} find that the elemental abundances of red clump stars belonging to the thin disk can be predicted almost perfectly from their age and [Fe/H] abundance. This implies that the key chemical evolution parameters affecting the elemental abundances (SSP parameters and yield tables) are held fixed, whilst ISM parameters vary smoothly over the thin disk \citep[which offsets the metallicity for different galactocentric radii, e.g.][for a simulated example]{Buck2020, Wang2024}. Similarly \cite{2019ApJ...874..102W} find that ISM parameter variations are deprojected in the [X/Mg] vs [Mg/H] plane (their Fig.\,17) and that abundance tracks in that space are independent of the stellar sample's spatial position within the Galaxy (their Fig.\,3).

Following \citet{Philcox_2019}, to avoid unrealistic star formation histories (that are very `bursty' for early stars), we additionally require that the SFR (parametrized by a $\Gamma$ distribution with shape parameter $a=2$\footnote{\texttt{CHEMPY} parametrizes the SFR with a $\Gamma$ distribution
\begin{equation*}
\label{eq:SFR}
\mathrm{SFR}\left(t,k,\vartheta\right)=\frac{1}{\Gamma(k)\vartheta^k}t^{k-1}\exp\left(\frac{-t}{\vartheta}\right), \mathrm{for}\,k=2 \rightarrow \vartheta = \mathrm{SFR}_\mathrm{peak} 
\end{equation*}
where the shape parameter is fixed to $k=2$ such that the scale parameter ($\vartheta$) determines the peak of the SFR}) at the maximum possible stellar birth-time ($13.8$\,Gyr) should be at least 5\% of the mean SFR, ensuring that there is still a reasonable chance of forming a star at this time-step. In our formalism, this corresponds to the constraint $\log_{10}\left(\mathrm{SFR}_\mathrm{peak}\right)>0.294$. For this reason, a truncated Gaussian prior will be used for the SFR parameter. Furthermore, we constrain $T_i$ to the interval $[1,13.8]$\,Gyr (assuming an age of the Universe of 13.8\,Gyr), ignoring any stars formed before $1$\,Gyr, which is justified as these are expected to be rare.

\paragraph{Nucleosynthetic yield tables:}
We adopt the same nucleosynthetic yield tables as in \citep{Philcox_2019}, see their Sec.~2.2 for more details.
To test our method, we aim further at inferring parameters from a sample of stars taken from a hydrodynamical simulation of a MW type galaxy which we take from the IllustrisTNG project \citep{Pillepich2018}. To ensure maximal compatibility with TNG, we adopt their nucleosynthetic yield tables in \texttt{CHEMPY}, for enrichment by SN\,Ia, SN\,II and AGB stars. The utilized yields are summarized in Tab.\,\ref{tab:chempy_TNG_yields}, matching \citet[Tab.\,2]{2018MNRAS.473.4077P}, and we note that the SN\,II yields are renormalized such that the IMF-weighted yield ratios at each metallicity are equal to those from the \citet{2006ApJ...653.1145K} mass range models alone. \texttt{CHEMPY} uses only net yields, such that they provide only newly synthesized material, with the remainder coming from the initial SSP composition. These tables may not well-represent true stellar chemistry, and the effects of this mismatch are examined in Sec.\,\ref{subsec:mocks_wrong_yield} by performing inference using an alternative set of yields that does not match the yield set of the training data. For the analysis of observational data, we would want to use the most up-to-date yields, such as \citet{2016ApJ...825...26K} AGB yields, and carefully choose elements which are known to be well reproduced by our current models (e.g. shown by \citet{2019ApJ...874..102W,2019arXiv190806113G}), though this is not appropriate in our context. To facilitate best comparison with Ilustris TNG, we further set the maximum SN\,II mass as $100\,\mathrm{M}_\odot$ (matching the IMF upper mass limit), adopt stellar lifetimes from \citet{portinari} and do not allow for any `hypernovae' \citep[in contrary to][]{2018ApJ...861...40P}).

\begin{table}[]
\caption{Nucleosynthetic yield tables used in this analysis, matching those of the TNG simulation \citep[Tab.\,2]{2018MNRAS.473.4077P}.}
     \centering
     \begin{tabular}{c|c}
       Type & Yield Table \\
        \hline
         SN\,Ia & \citet{1997NuPhA.621..467N}\\
         SN\,II & \citet{2006ApJ...653.1145K,portinari}\\
         AGB & \citet{2010MNRAS.403.1413K,2014MNRAS.437..195D};\\
         & \citet{2014ApJ...797...44F}
     \end{tabular}
 \label{tab:chempy_TNG_yields}
 \end{table}

\paragraph{Chemical elements:}
In our analysis we only track nine elements: C, Fe, H, He, Mg, N, Ne, O and Si since these are the only elements traced by TNG. We principally compare the logarithmic abundances [X/Fe] and [Fe/H] defined by 
\begin{equation}
     [\mathrm{X}/\mathrm{Y}] = \log_{10}(N_\mathrm X/N_\mathrm Y)_\mathrm{star} - \log_{10}(N_\mathrm X/N_\mathrm Y)_\odot
\end{equation}
for number fraction $N_\mathrm X$ of element X. Here $\odot$ denotes the solar number fractions of \citet{2009ARA&A..47..481A}. As is customary, we use H for normalization, thus we are left with $n_\mathrm{el}=8$ independent elements which must be tracked by \texttt{CHEMPY}\footnote{In observational contexts, it may be more appropriate to compute abundances relative to Mg rather than Fe, as in \citep{2019ApJ...874..102W}, since Mg is only significantly produced by SN\,II and hence a simpler tracer of chemical enrichment.}. 

With these modifications, \texttt{CHEMPY} allows to predict TNG-like chemical abundances for a given set of galactic parameters. It is important to note that the two GCE models have very different parametrizations of galactic physics, with TNG including vastly more effects, it is thus not certain \textit{a priori} how useful \texttt{CHEMPY} will be in emulating the TNG simulation, although its utility was partially demonstrated in \citet{2018ApJ...861...40P}. However,  such a test is necessary to prepare for an inference on real data.

\subsection{Neural network emulator for \texttt{CHEMPY}}
Despite the simplifications made by the GCE model \texttt{CHEMPY}, the run-time of \texttt{CHEMPY} and the high-dimensionality of the parameter space incurs some difficulties when sampling the distribution of the global parameters $\vec\Lambda = \{\alpha_\mathrm{IMF},\log_{10}(N_\mathrm{Ia})\}$. 
To alleviate this, we follow \cite{Philcox_2019} and implement a \textit{neural network} (NN) emulator of the \texttt{CHEMPY} simulator. We design the NN as a simple feed-forward neural network with 2 hidden layers and 100 neurons in the first and 40 neurons in the second layer. The NN is trained on $\sim700,000$ data points and validated on $\sim50,000$ additional data points created with \texttt{CHEMPY} using a uniform prior over the $5\sigma$-range of the original Gaussian prior stated in table~\ref{tab:priors}. The batch size is set to 64 and the learning rate is set to 0.001. We train for 20 epochs using a schedule free optimizer \citep{schedulefree}. 
Training this tiny emulator takes about 200s on the CPU.

In essence, instead of computing the full model for each input parameter set, we pass the parameters to the NN which predicts the output abundances to high accuracy. As already argued in \cite{Philcox_2019} this has two benefits;
\begin{enumerate}
    \item \textbf{Speed:} The run-time of the \texttt{CHEMPY} function is $\sim1$\,s per input parameter set, which leads to very slow generation of training data for SBI. With the NN emulator, this reduces to $\sim5\times10^{-5}$\,s, and is trivially parallelizable, unlike \texttt{CHEMPY}. Having access to a fast simulator opens the possibility for testing Sequential Neural Posterior Estimate (SNPE) as an alternative, but we discuss this possibility in Section \ref{sec: discussion}. 
    \item \textbf{Differentiability:} The NN is written in pytorch and has additionally a simple closed-form analytic structure \citep[described in the appendix of][]{Philcox_2019}, unlike the complex \texttt{CHEMPY} model. Both aspects allow it to be differentiated either via auto-diff or analytically, so one can use it to sample via advanced methods such HMC as done in \citep{Philcox_2019}.
\end{enumerate}

\begin{figure}[]
     \centering
     \includegraphics[width=\columnwidth]{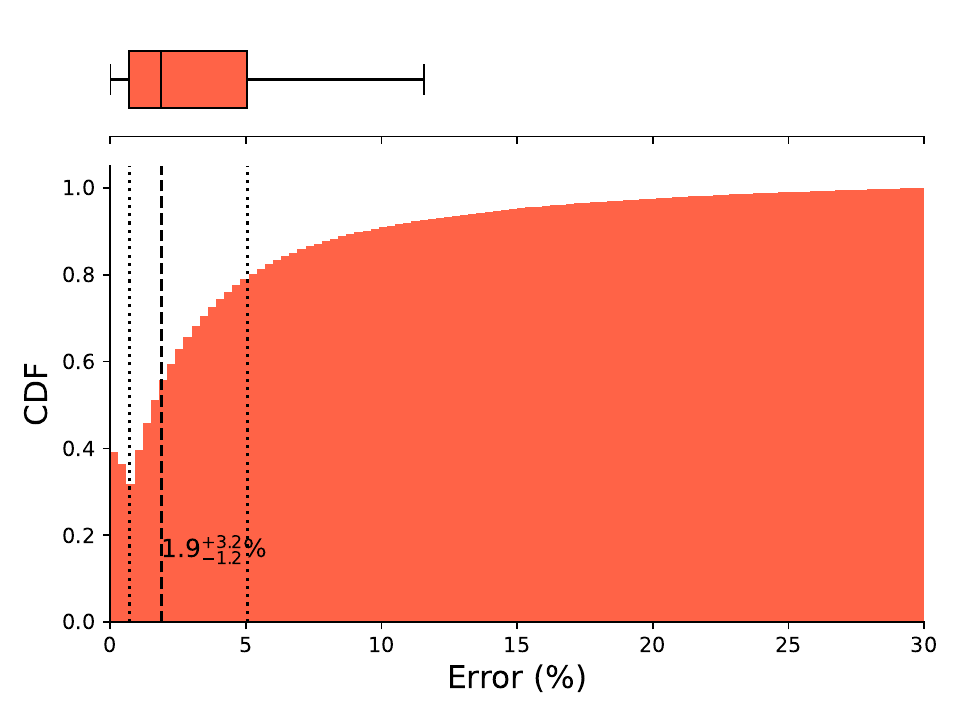}
     \vspace{-.5cm}
     \caption{Cumulative absolute percentage error of the NN emulator for the \texttt{CHEMPY} simulator. The orange histogram shows the cumulative distribution of percentage errors with the vertical dashed line indicating the median and the vertical dotted lines indicating the first and third quartile. The box plot on the top of the plot extends from the first quartile to the third quartile of the data, with a line at the median. The whiskers extend from the box to the farthest data point lying within $1.5\times$ the inter-quartile range from the box. The NN predicts abundances with an absolute percentage error far below typical observational errors.}
     \label{fig:ape_NN}
     \script{evaluate_emulator.py}
\end{figure}

Despite the additional complexity introduced by using multiple stellar data-points, our NN simply needs to predict the birth-time abundances for a single star (with index $i$) from a given set of six parameters; $\{\vec\Lambda,\vec\Theta_i,T_i\}$. The same NN can be used for all $n_\mathrm{stars}$ stars (and run in parallel), reducing a set of $n_\mathrm{stars}$ runs of \texttt{CHEMPY} to a single matrix computation. With the above network parameter choices, the NN predicts abundances with an absolute percentage error of %
  $1.9^{+3.2}_{-1.2}\,\%$%
\unskip\label{output/ape_NN.txt}\unskip%
 (which translates into a logarithmic error of %
  $0.008\,$%
\unskip\label{output/ape_NN_log.txt}\unskip%
 dex)  
which is far below typical observational errors and even smaller away from the extremes of parameter space (see Fig.~\ref{fig:ape_NN})\footnote{Our NN emulator is publicly available on the github repository accompanying this manuscript \href{https://github.com/TobiBu/sbi-chempy/blob/main/src/scripts/train_torch_chempy.py}{https://github.com/TobiBu/sbi-chempy/blob/main/src/scripts/train\_torch\_chempy.py} with pre-trained NN weights available on zenodo \href{https://zenodo.org/records/14925307}{https://zenodo.org/records/14925307}.}. In fact, we will add additional observational uncertainty to our mock observational data later during training of the neural posterior estimator network.

\subsection{Bayesian model}
\texttt{CHEMPY} effectively is a Bayesian model for stellar abundances given a set of parameters $\{\Vec{\Lambda},\Vec{\Theta},T\}$ and \citet{Philcox_2019} extended the \texttt{CHEMPY} framework to be able to model multiple stellar data-points. Consider a given star with index $i$ that is born in some region of the ISM. This star will carry its own set of parameters $\{\Vec{\Lambda},\Vec{\Theta}_i,T_i\}$, where $\Vec{\Lambda}$ are star independent and hence taken to be global parameters while the ISM parameters $\Vec{\Theta}_i$ and the birth-time $T_i$ are star specific. Using \texttt{CHEMPY} (or the trained neural network emulator) we can easily model the set of $n_\mathrm{el}$ chemical abundances $\{X_i^j\}$ for the $i$-th star as: 
\begin{equation}\label{eq:chempy_function}
\{X_i^j\} = \texttt{CHEMPY}(\Vec{\Lambda},\Vec{\Theta}_i,T_i),
\end{equation}
where $j$ indexes the chemical element. 
These model abundances can then be compared against observations, with measured abundances $d_i^j$ and corresponding Gaussian errors $\sigma_{i,\mathrm{obs}}^j$ jointly denoted as $D_i=\{d_i^j,\sigma_{i,\mathrm{obs}}^j\}$.

\paragraph{Posteriors for Galactic parameters}
As stated in \citet{Philcox_2019} the full posterior for this case is given by 
\begin{eqnarray}\label{eq:posterior}
    \mathbb{P}(\vec\Lambda,\{\vec\Theta_i\},\{T_i\}|\{D_i\}) &\propto&  \left[\prod_{i=1}^{n_\mathrm{star}}p_{\vec\Theta}(\vec\Theta_i)p_{T_i}(T_i)\right]
    \times p_{\vec\Lambda}(\vec\Lambda)\\
    \nonumber
    &\times& \mathcal{L}(\{D_i\}|\vec\Lambda,\{\vec\Theta_i\},\{T_i\})
\end{eqnarray}
where $p_{\vec\Theta}(\vec\Theta_i)p_{T_i}(T_i)$ are the priors on the variables $\vec\Theta_i$ or $T_i$ belonging to a given set of stars.

In order to determine the optimal values of the global galactic parameters ($\Vec{\Lambda}$) one has to sample the posterior of Eq.~\ref{eq:posterior}. In practice this is a costly computation, since even with advanced techniques such as HMC sampling the posterior can only be evaluated for a small set of stars ($\lesssim200$) and requires long compute times \citep[$\sim42$ hours][]{Philcox_2019}.
However, recent advances in implicit-likelihood inference or SBI \citep{Cranmer2020} offer another very efficient approach to approximate the posterior (see next paragraph for more details). These methods train a neural conditional density estimator to represent the conditional posterior, $\mathbb{P}(\vec\Lambda|\{D_i\})$, which can be very efficiently evaluated given observational data $\{D_i\}$.

In particular, if we marginalize over the star specific parameters and solely focus on the global parameters $\Vec{\Lambda}$ we can make the assumption that individual observations of stars are identically and independently distributed (i.i.d.) and factorize the joint posterior from above to simply express it as:  %
\begin{eqnarray}
\mathbb{P}(\vec\Lambda|\{D_i\}) &\propto& 
\mathbb{P}(\vec\Lambda)\mathbb{P}(\{D_i\}|\Lambda)\quad\quad\quad\quad\quad\quad\,\, \text{(Bayes rule)}
\\\nonumber
&=&
\mathbb{P}(\vec\Lambda)\mathbb{P}
(D_1,...,D_{n_\mathrm{star}}|\vec\Lambda)
\\\nonumber
&=& \mathbb{P}(\vec\Lambda)\prod_{j=1}^{n_\mathrm{star}}\mathbb{P}(D_j|\vec\Lambda) \quad\quad\quad\quad\quad\quad\quad\,\,\,\, \text{(i.i.d.)}
\\\nonumber
&\propto& \mathbb{P}(\vec\Lambda) \prod_{j=1}^{n_\mathrm{star}}\frac{\mathbb{P}(\vec\Lambda|D_j)}{\mathbb{P}(\vec\Lambda)} \quad\quad\quad\quad\quad\text{(Bayes rule)}
\\
\label{eq:posterior}
&=& \mathbb{P}(\vec\Lambda)^{1-n_\mathrm{star}}\ \prod_{j=1}^{n_\mathrm{star}}\mathbb{P}(\vec\Lambda|D_j) 
\end{eqnarray}

This factorization of the joint posterior into single star posteriors $\mathbb{P}(\vec\Lambda|D_j)$ has the advantage that we do not need to specify the exact number of observational data points beforehand. We can simply train our neural density estimator on single star observations and then at evaluation time, we simply combine as many posterior estimates as we have observational data.

\paragraph{Simulation-based inference:}
In a nutshell, SBI \citep[e.g.][]{Cranmer2020,Papamakarios:2021,Gloeckler2024AllinoneSI} -- also called likelihood free inference within a Bayesian inference framework -- works as follows: given an assumed generative model $\mathcal{M}$ of parameters $\Vec{\theta}$ (in our case a GCE model) and a set of simulated observations of individual stellar abundances $\Vec{X}$ from that model, we train a mapping between the two to estimate the posterior distribution $p$($\Vec{\theta}|\Vec{X}$, $\mathcal{M}$) of the generative model parameters $\Vec{\theta}$ that reproduce the simulated observations $\Vec{X}$. Once this mapping is trained, we can apply it to observations of stellar abundances $\Vec{X_R}$ to infer $p$($\Vec{\theta}|\Vec{X_R}$, $\mathcal{M}$). We show a schematic visual representation of our method in Fig. \ref{fig:flowchart}. Note, we do not need to know anything about $p$($\Vec{\theta}|\Vec{X_R}$, $\mathcal{M}$), we solely need to be able to sample from it.

We use a NPE \citep{zeghal2022neuralposteriorestimationdifferentiable} which utilises the gradients of the generative model $\mathcal{M}$ with a Masked Autoregressive Flow 
\citep[MAF;][]{papamakarios2018maskedautoregressiveflowdensity} 
containing 8 hidden features and 4 transformation layers for the normalizing flow. 
The expressivity of the MAF allows the NDE to be capture complex distributions, while also maintaining computational tractability. The final model was selected after extensive hyperparameter tuning, varying: the architecture between Neural Spline Flow \citep[NSF;][]{durkan2019neuralsplineflows}, MAF, and MAF with rational-quadratic spline (MAF-RQS), the number of neurons between {10, 20, 50, 100}, and the number of transformations between {1, 5, 10, 30}, optimizing over the test set for both the highest mean log posterior probability and the best calibration as measured by the TARP value \citep{Lemos2023}. For more details on the TARP value see section \ref{sec:sbc} in the appendix.

We train our NPE with $10^5$ data points consisting of $n_{\rm elements}=8$ simulated with the NN emulator described in the previous section. Inputs are sampled from the Gaussian priors shown in Tab. \ref{tab:priors}. Training takes $\sim10$ minutes on an Apple M1 Max.
We evaluate the accuracy of the NPE using $5\times10^3$ validation data points from the original \texttt{CHEMPY} simulator. In order to mimic observational noise, we add 5\% observational uncertainties to the abundances simulated with \texttt{CHEMPY} before feeding them to our NPE.
We have made sure that our NPE is well calibrated and posterior distributions are trustable. For more details on simulation-based calibration see appendix \ref{sec:sbc}.

Note, the methods presented here can naturally be extended to any fast and flexible GCE model \citep[e.g.][and others]{Talbot1971,vice,omega2018}, not just \texttt{CHEMPY} and can even be used to infer vastly different galactic parameters such as accretion events from abundance distributions of stars \citep[e.g.][]{Viterbo2024}.

\paragraph{Multi-star inference:}
Following eq.~\ref{eq:posterior} we can calculate the joint posterior for a combined inference using abundance observations of multiple stars. For this, we will condition our NPE individually on single star observations and sample the posterior. Then, according to eq.~\ref{eq:posterior} we simply have to multiply individual posteriors and account for the inverse weighting by some power of the prior which we take from Table~\ref{tab:priors}. 

This does however present one issue, that since we only have access to samples from the posterior and not the posterior itself it is difficult to evaluate eq.~\ref{eq:posterior}. We circumvent this by approximating each single star posterior by a Gaussian and fit for the parameters of mean and covariance. With this it is straight forward to evaluate eq.~\ref{eq:posterior} analytically. In fact, under our assumption the combined posterior is a product between the prior and the product of multiple Gaussians for the individual star posteriors. The latter product is also a Gaussian with mean $\mathbf{\mu'}$ and variance $\mathbf{\sigma'^2}$:
\begin{align}
    \mathbf{\mu'} &= \frac{\sum_{i=1}^{N_{stars}} \frac{\mu_i}{\sigma_i^2}}{\sum_{i=1}^{N_{stars}} \frac1{\sigma_i^2}} \\
\mathbf{\sigma'}^2 &= \frac1 {\sum_{i=1}^{N_{stars}} \frac1{\sigma_i^2}}
\end{align}

Further, in our case the prior for the galactic parameters $\Lambda$ is Gaussian as well. Therefore the resulting factorized posterior from eq.~\ref{eq:posterior} is again a Gaussian and can be expressed with mean $\mathbf{\mu}$ and variance $\mathbf{\sigma}$ as:
\begin{align}
\mathbf{\mu} &= \frac{\frac{\mu'}{\sigma'^2}-\frac{(1-N)\mu_ {prior}}{\sigma_ {prior}^2}}{\frac1{\sigma'^2}-\frac{(1-N)}{\sigma_ {prior}^2}} \\
\mathbf{\sigma}^2 &= \frac1 {\frac1{\sigma'^2}-\frac{(1-N)}{\sigma_ {prior}^2}}
\end{align}

Given the tiny and simple neural network that represents our NPE, we note that the above assumption of Gaussianity in each of the single star posteriors not expected to notably increase our pipeline's error. We have further empirically verified that single star posteriors are indeed close to Gaussian.
However, in future work we plan to alleviate this simplification and directly approximate the joint posterior of a multi-star inference.

\section{Results}
\label{sec: Results}

\begin{figure}[]
     \centering
     \includegraphics[width=\columnwidth]{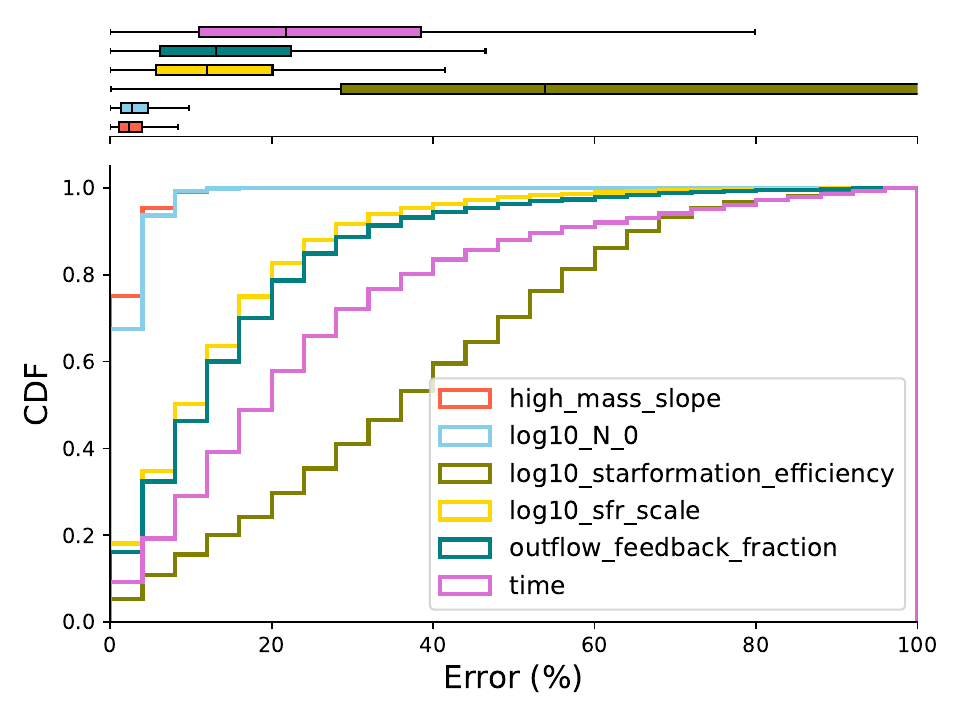}
     \vspace{-.5cm}
     \caption{Absolute percentage error of the neural posterior density estimate for a single star. Different colored histograms show the full error distribution for all 6 parameters of interest with the median values highlighted by the vertical dashed lines. The box plots show again the first and third quantiles with the median represented by a vertical line and the whiskers extending from the box to the farthest data point lying within $1.5\times$ the inter-quartile range from the box. The global parameters of main interest for this work are shown by the light blue and red histogram.}
     \label{fig:posterior_APE}
     \script{plot_posterior_APE.py}
\end{figure}

We use our SBI method described in the previous section to infer the global galactic parameters $\vec\Lambda = \{\alpha_\mathrm{IMF},\log_{10}(N_\mathrm{Ia})\}$. In order to demonstrate the performance and robustness of our methods we use three mock data-sets: 
\begin{enumerate}
    \item Mock observations drawn from \texttt{CHEMPY} from the same yield set as the training data for the neural network emulator. With this we ensure to test our training strategy and the performance without any systematic distribution shifts. 
    \item  \texttt{CHEMPY} mock data using a different yield set to test for potential biases through model misspecification in our SBI setup. (see  \ref{tab:chempy_ALT_yields})
    \item Simulated data from a large-scale hydrodynamical simulation taken from the IllustrisTNG suite \citep{Pillepich2018} but with the same yield set as our \texttt{CHEMPY} training data to ensure that we recover the correct parameters even for models with a completely different treatment of ISM physics.
\end{enumerate}

\subsection{\texttt{CHEMPY} Mock Observational Data}
\label{subsec:mock obs}
Our analysis uses mock observations drawn from the neural network emulator with fixed values of the global galactic parameters $\alpha_{\rm IMF}=-2.3$ and $\log_{10}(N_{\rm Ia})=-2.89$ and varying local parameters $\vec{\Theta}_i$ using priors from Tab. \ref{tab:priors}. Additionally we draw $T_i$ uniformly in the range $[2,12.8]$ Gyr to minimize overlap with the neural network training birth-time limits when observational uncertainties are included. 
Each set of parameters is passed to our \texttt{CHEMPY} emulator, producing eight true chemical abundances. In order to fully represent observational data, we augment this data with observational uncertainties by adding a Gaussian error of $0.05$ dex for the abundances representative for typical APOGEE data \citep{Majewski2016}.

\begin{figure*}
    \centering
    \includegraphics[width=\textwidth]{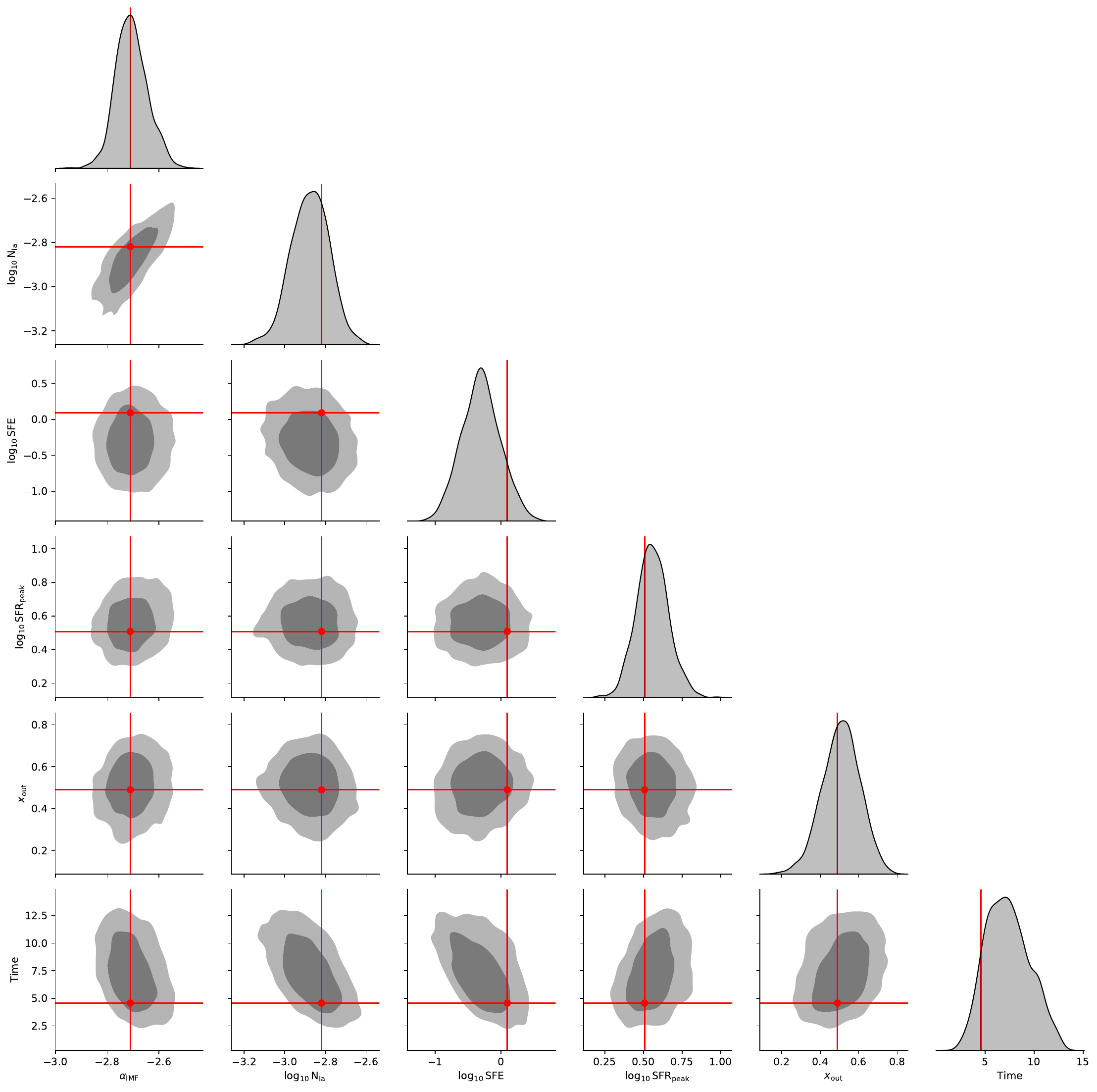}
    \vspace{-.5cm}
    \caption{Corner plot of the posteriors for all six parameters for a single star from the validation set. The gray contours show a kde-estimate of the posterior from our SBI inference and the red dot and lines show the ground truth parameter values. Gray histograms on the diagonal show a kde estimate of the marginals.}
    \label{fig:corner_plot}
    \script{plot_sbc.py}
\end{figure*}

\begin{figure*}
    \centering
    \includegraphics[width=\textwidth]{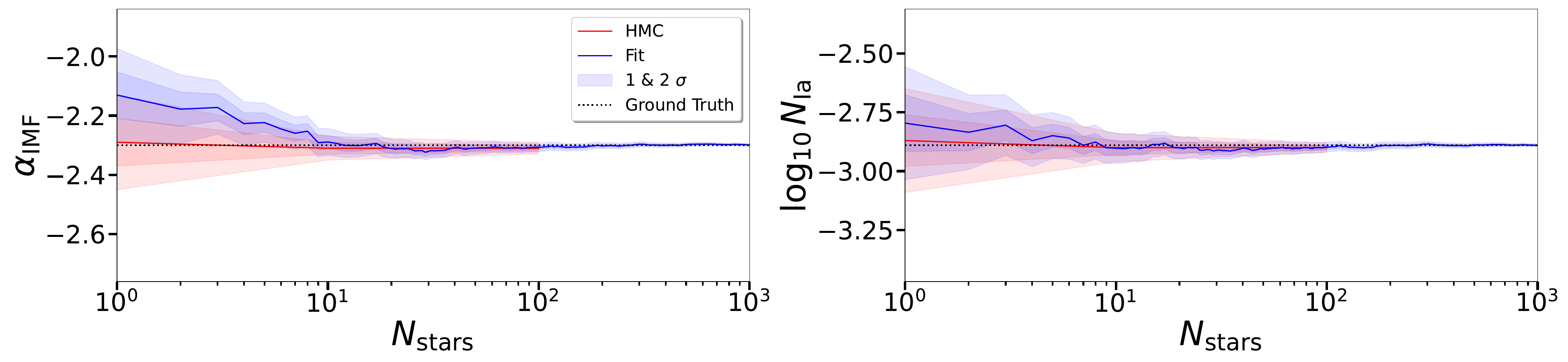}
    \vspace{-.5cm}
    \caption{Accuracy of inferred global galactic parameters $\alpha_{IMF}$ and $\log_{10}(N_{Ia})$ as a function of number of observed stars. We compare our SBI results (blue line) to the inferred values using HMC (red line) as done by \cite{Philcox_2019} and the ground truth values (black dashed line) for various test cases as described in Sec. For the SBI analysis we show $1\sigma$ and $2\sigma$ contours while HMC results only show $1\sigma$ statistical uncertainties as reported in Tab.~3 of \citet{Philcox_2019} (blue/red shaded regions). See Sec.~\ref{subsec:chempy_tng} for a full description.}
    \label{fig:CHEMPY_TNG_N_star_analysis}
    \script{chempy_tng_inference.py}
\end{figure*}

\begin{figure}
    \centering
    \includegraphics[width=\columnwidth]{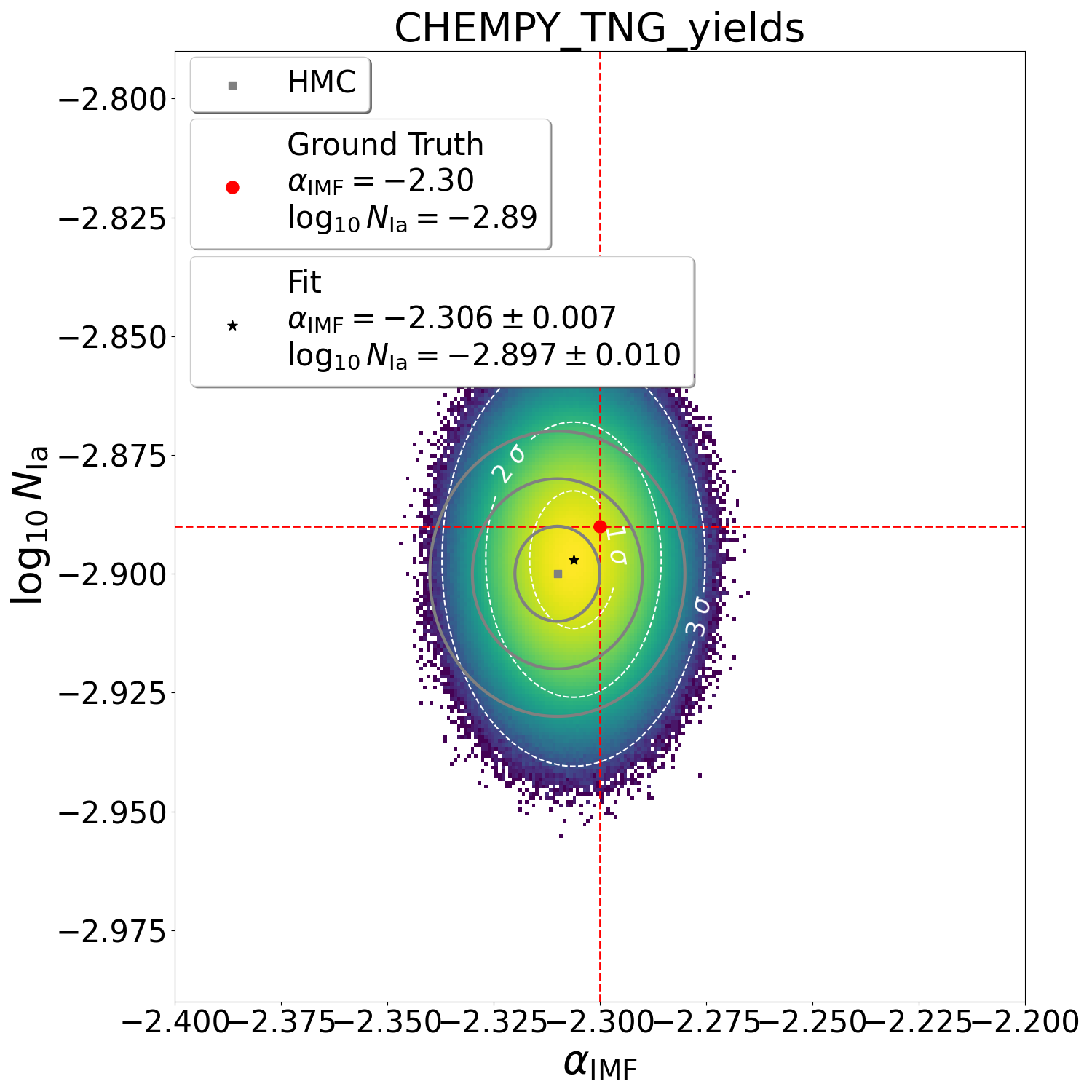}
    \vspace{-.25cm}
    \caption{Joint posterior for global galactic parameters $\alpha_{\rm IMF}$ and $\log_{10}(\rm N_{Ia})$ of $100$ stars. The ground truth value is shown by the red star, the posterior mean of the SBI inference is shown with the black dot and white ellipses show the $1-3\sigma$ contours of our inference. The gray ellipses show the results for HMC inference performed by \citet{Philcox_2019}. See Sec.~\ref{subsec:chempy_tng} for a full description.}
    \label{fig:CHEMPY_TNG_sbi} 
    \script{chempy_tng_inference.py}
\end{figure}
\begin{figure*}
    \centering
    \includegraphics[width=\textwidth]{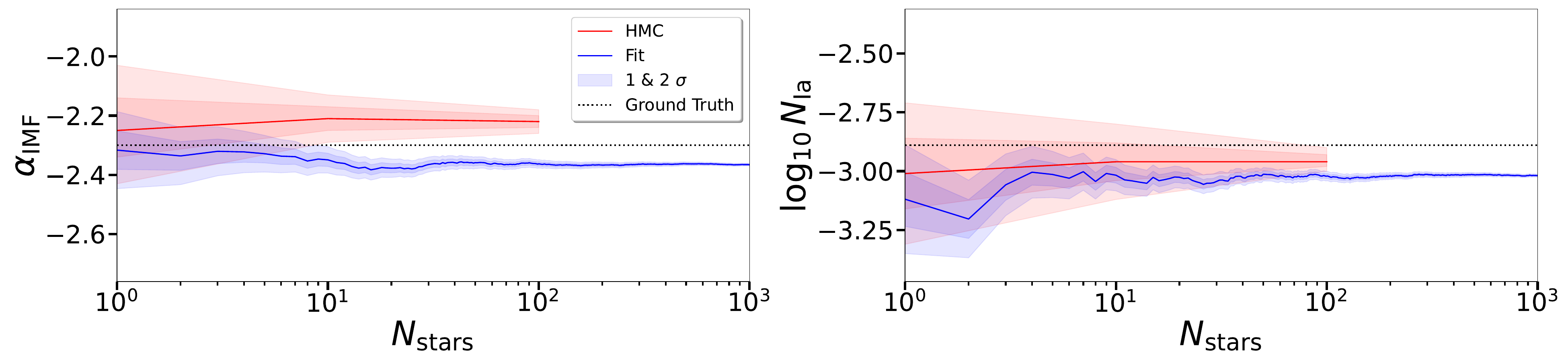}
    \vspace{-.5cm}
    \caption{Same as Fig.~\ref{fig:CHEMPY_TNG_N_star_analysis} but for the mock data created with a different yield set than the training data. See Sec.~\ref{subsec:mocks_wrong_yield} for a full description.}
    \label{fig:CHEMPY_alt_N_star_analysis}
    \script{chempy_alt_yield_inference.py}
\end{figure*}
\begin{figure*}
    \centering
    \includegraphics[width=\textwidth]{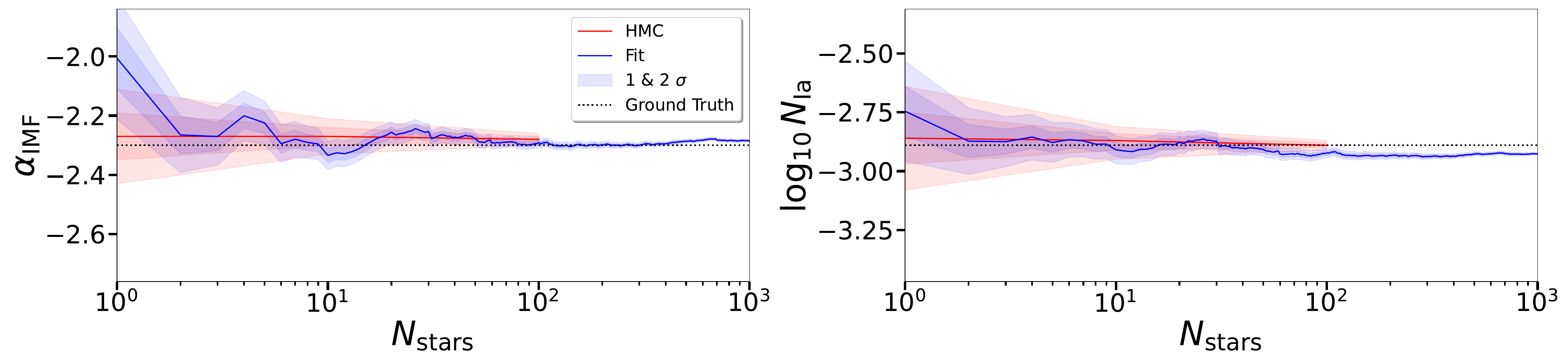}
    \vspace{-.5cm}
    \caption{Same as Fig.~\ref{fig:CHEMPY_TNG_N_star_analysis} but for the mock data taken from an IllustrisTNG Milky Way-like galaxy. See Sec.~\ref{subsec:tng_sim} for a full description.}
    \label{fig:TNG_N_star_analysis}
    \script{tng_inference.py}
\end{figure*}

For each individual observation consisting of the abundance measurements of a single star we sample the posterior estimate for all six parameters $\{\vec\Lambda,\vec\Theta_i,T_i\}$ with 1,000 points. This takes around $0.3s$ for each star, making it extremely fast to infer the parameters of a large amount of stars. Our method takes around $10$ seconds to build a posterior for all six parameters for a dataset size of 1,000 stars (each time sampling the single star posterior with 1,000 points and fitting for the Gaussian parameters). Our combined runtime for the NPE training plus sampling is then $610$ seconds on an Apple M1 Max for 1,000 stars hence our methods making it more than $1240$ times faster than current HMC methods which take around $42$h for only $200$ stars \citep{Philcox_2019}. Note though, that since our approach is amortized and we do not need to retrain our NPE model each time we want to make an inference, we in fact achieve an inference speed-up of a factor of $\sim75,600$.
Hence, shorter computing times make it feasible to use orders of magnitudes more observations.

\paragraph{Validation through absolute percentage error:}
We start evaluating our method by using \texttt{CHEMPY} to produce a mock observational dataset to ensure no systematical shift between training and testing data in terms of physics parameters.

To evaluate our method quantitatively, we compare the posterior mean to the ground truth value for each observation and calculate the absolute percentage error (APE, see Fig.~\ref{fig:posterior_APE}). For a single star observation, our NPE has an APE of %
  $9.3^{+17.1}_{-6.3}\,\%$%
\unskip\label{output/global_posterior_APE.txt}\unskip%
 when looking at all 6 parameters of interest.
If we restrict ourselves to only the two global parameters $\vec\Lambda$, our NPE achieves an APE of %
  $2.3^{+1.7}_{-1.2}\,\%$
\unskip\label{output/posterior_APE.txt}\unskip%
 as shown in Fig.~\ref{fig:posterior_APE}. There we can also see, that the accuracy for an individual star of the NPE is not particularly high. 

However, currently our NPE network is not particularly good at estimating ages from abundances alone. On average our inference for ages results in an APE of $\sim21\%$ which is slightly larger than the observational noise of $20\%$ that we add during the mock up of our data but well in agreement with current uncertainties of stellar age inference that range from 15\% to 30\% for turn-off stars and giants respectively.

We accompany the APE analysis by showing the full posterior inference results for a single star in Fig.~\ref{fig:corner_plot}. This figure shows that the SBI approach is well able to infer correct parameters and their cross-correlations. In partciular, there is a strong correlation between the two global parameters as seen in the top left corner as well as for time (or stellar age) and all other five parameters as visible from the bottom row.

Finally, as already mentioned in the method section for the global parameters $\vec\Lambda$ we can boost the accuracy by combining the inference for many stars of the same galaxy.

\subsection{Inference on mock data from \texttt{CHEMPY} with TNG yield set}
\label{subsec:chempy_tng}
Combining the inference on multiple observed stars gives us higher accuracy and precision of the global galactic parameters $\vec\Lambda$. We therefore perform inference using a range of stars $n\in[1,1000]$ to show how inference accuracy increases with number of observations (see Fig. \ref{fig:CHEMPY_TNG_N_star_analysis}). We find that in the limit of less than $\sim100$ stars SBI shows a larger uncertainties than the HMC results of \citet{Philcox_2019}. But when using more than a few hundred stars the accuracy and precision is superior compared to HMC. In particular, after using a few tens of stars, the NPE estimate is already less biased than the HMC results. 
Finally, given our computational advantage we will be able to use orders of magnitude more stars for our inference.
This is particularly important since sample variances play a large role when using small samples of stars as also noted in \citet{Philcox_2019}.

In Fig.~\ref{fig:CHEMPY_TNG_sbi} we show the joint posterior for $\alpha_{IMF}$ and $\log_{10}(N_{Ia})$ for our inference. The red star indicates the ground truth value, the black dot shows our posterior mean and white contours indicate $1-3\sigma$ levels.
Using a sample of 1,000 stars we infer %
  $\alpha_{\rm IMF}=-2.299\pm0.002$ and $\log_{10}(\rm N_{Ia})=-2.890 \pm 0.003$%
\unskip\label{output/CHEMPY_TNG_sbi.txt}\unskip%
 which deviates less than $\sim0.04\%$ from the ground truth value. We have further checked the accuracy of our inference for a vastly different mock observational dataset created with shifted parameters of $\alpha_{IMF}=-2.1$ and $\log_{10}(N_{Ia})=-3$ and found that also in this case our model is well able to lead to correct inferences (see sec.~\ref{sec: additional inference} in the appendix).

Note that our analysis is in principle also able to infer the local parameters $\Vec{\Theta}_i$ and $T_i$. This would allow us to estimate/infer stellar ages as well. But note, as discussed above at the end of sub-section \ref{subsec:mock obs} our NPE currently is not well calibrated to estimate stellar ages accurately enough.

In summary, our SBI pipeline is quite capable of correctly and precisely inferring  global parameters of chemical enrichment models from stellar abundance alone when using the same physical model and yield tables as during training. Next, we will check what happens if the training data is generated with a different yield set to that use at inference time.

\subsection{Inference on mock \texttt{CHEMPY} data with incorrect yield set}
\label{subsec:mocks_wrong_yield}

\begin{figure}
    \centering
    \includegraphics[width=\columnwidth]{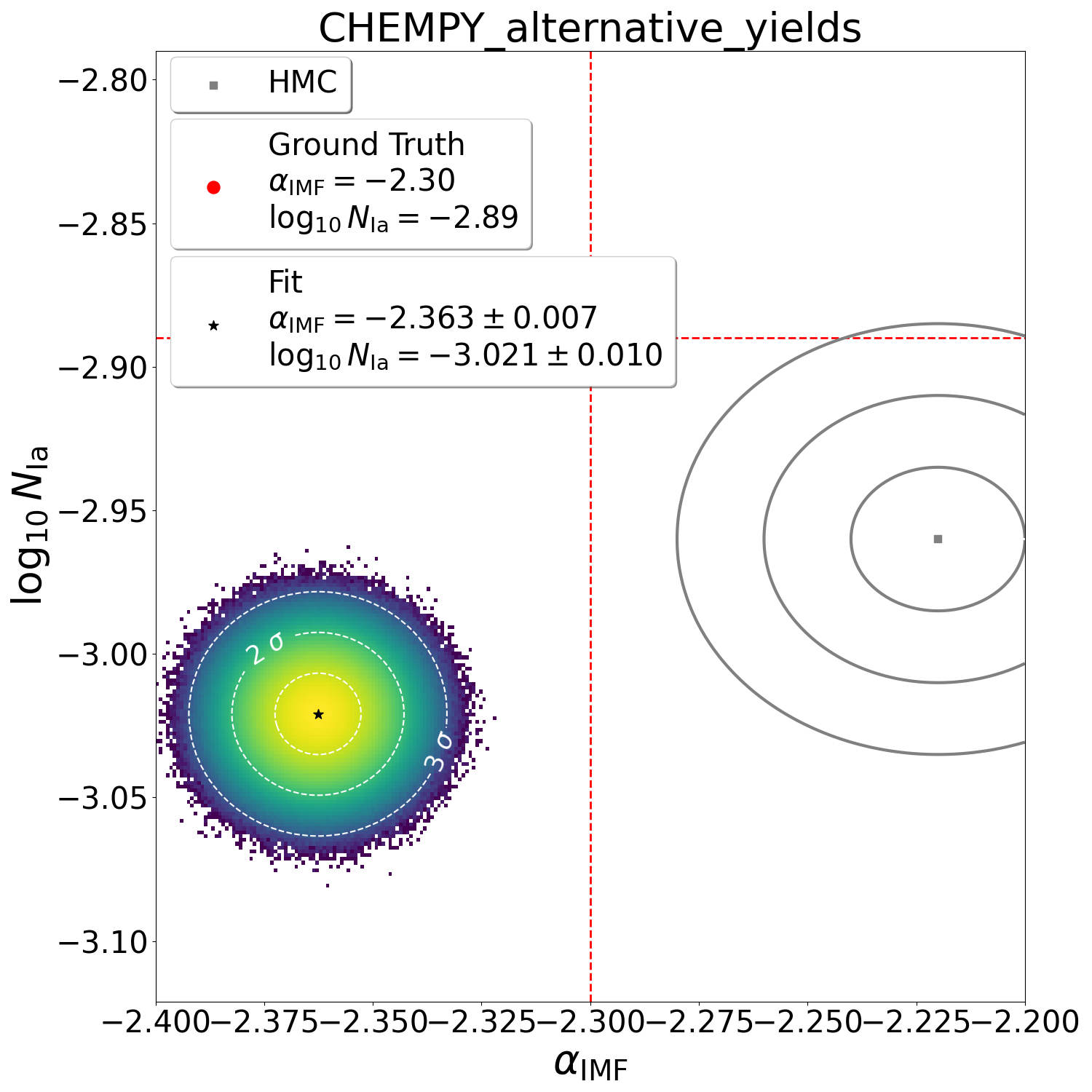}
    \vspace{-.25cm}
    \caption{Same as Fig.~\ref{fig:CHEMPY_TNG_sbi} but for the mock data created with a different yield set than the training data. See Sec.~\ref{subsec:mocks_wrong_yield} for a full description.}
    \label{fig:CHEMPY_alt_sbi} 
    \script{chempy_alt_yield_inference.py}
\end{figure}

There is an extensive discussion in the literature about stellar nucleosynthesis with various different yield sets proposed \citep[see e.g. the discussion in][]{Rybizki_2017}. In fact, all tabulated yield sets currently differ from reality and during an application of our inference pipeline it will not be clear which tabulated yield set most closely matches reality and hence which should be used.
In order to investigate how sensitive our method is to model misspecification by using an incorrect yield set, we create another set of mock data using \texttt{CHEMPY} with a different yield set (Tab. \ref{tab:chempy_ALT_yields}) than during training of our NPE. For better cross-comparison we decided to use the same alternative yield sets as presented in Tab.~5 of \citep{Philcox_2019}. 

\begin{table}[]
\caption{Alternative nucleosynthetic yield tables used for model misspecification tests.}
     \centering
     \begin{tabular}{c|c}
       Type & Yield Table \\
        \hline
         SN\,Ia & \citet{2003NuPhA.718..139T} \\
         SN\,II & \citet{Nomoto2013} \\
         AGB & \citet{2016ApJ...825...26K}
     \end{tabular}
\label{tab:chempy_ALT_yields}
\end{table}

By choosing this set of yields we have made sure that contributions to all three processes differ by $\mathcal{O}(10\%)$. For more details see also Sec.~6.2 of \citet{Philcox_2019}.

\begin{figure}
    \centering
    \includegraphics[width=\columnwidth]{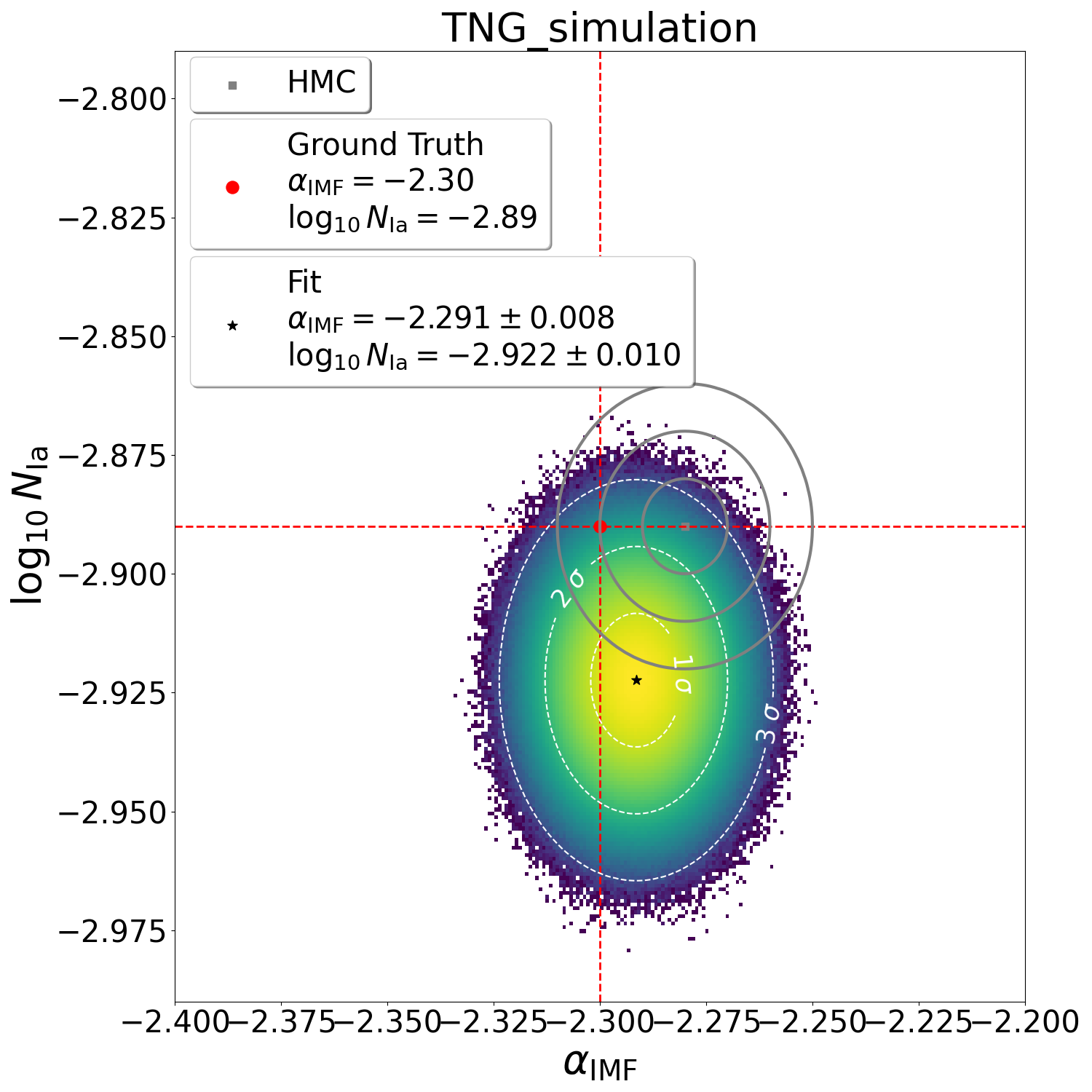}
    \vspace{-.25cm}
    \caption{Same as Fig.~\ref{fig:CHEMPY_TNG_sim_sbi} but for the mock data taken from an IllustrisTNG Milky Way-like galaxy. See Sec.~\ref{subsec:tng_sim} for a full description.}
    \label{fig:CHEMPY_TNG_sim_sbi} 
    \script{tng_inference.py}
\end{figure}

Our mock data generation and inference then follows the one of Sec.~\ref{subsec:chempy_tng}. This means we apply our NPE that we trained on \texttt{CHEMPY} stellar abundances simulated with the Ilustris TNG yield set to a sets of stellar abundances simulated with \texttt{CHEMPY} but using the alternative yield sets mentioned in Tab. \ref{tab:chempy_ALT_yields}. This effectively probes the effect of model misspecification on the inference results.

In Fig.~\ref{fig:CHEMPY_alt_N_star_analysis} we show our inference results (blue) for this setup for varying number of stars and compare them against HMC results (red) from \citet{Philcox_2019}. We see that the SBI results are similar biased as the HMC results. In fact with 100 stars HMC inferences are about 4 and 3 $\sigma$ away from the ground truth value for $\alpha_{\rm IMF}$ and $\log_{10}(N_{\rm Ia})$, respectively. Again, the joint posterior is shown in Fig.~\ref{fig:CHEMPY_alt_sbi} and shows that while individually parameter inferences look good, jointly taken the SBI results are on the edge of being $3\sigma$ away from the ground truth.

Nevertheless, the performance of our inference is still very good. We see that an increasing number of observations helps to decrease the models uncertainty just as before. However, we also note that our inference is now slightly biased as the observational data does not match the training data. Looking at Fig.~\ref{fig:CHEMPY_alt_sbi} we see that our inference is inconsistent with the ground truth within several $\sigma$ levels. Comparing this to the HMC results (red band in Fig.~\ref{fig:CHEMPY_alt_N_star_analysis}), we see that SBI is performing similarly bad as HMC when the model is misspecified. Part of the problem here is that through our assumption of a factorized posterior, we decrease the inference uncertainty as we increase the number of observations. In future we will improve upon this through different model architectures. In summary, the drastically reduced compute times offer a key advantage of our SBI method compared to more standard approaches such as HMC. Nevertheless, in the limit of large star counts our current approach becomes over-confident and model uncertainities are underestimated.

In an accompanying paper we more closely investigate measures of model misspecification and inference of best fitting models next to just parameter inference.  

\subsection{Inference on mock data from a IllustrisTNG simulated galaxy}
\label{subsec:tng_sim}

As a GCE code, \texttt{CHEMPY} is a one-zone model with simplified ISM physics that only approximately describes star formation and feedback as well as metal mixing in the ISM.
In the parametrization of \texttt{CHEMPY} as used here, we can assign each star to its own ISM environment, but we cannot exchange gas between environments and do not model sudden star formation or infall events. Hence, this section is dedicated to investigating whether this significantly biases our inference of the SSP parameters (noting that results from \citet{2019ApJ...874..102W} justify the treatment of ISM parameters as latent variables).

In order to explore what effect this simplified treatment of ISM physics has on the inference we now turn to a more complex model of the formation and chemical enrichment of a Milky Way-type galaxy taken from the IllustrisTNG simulations. Note, that by now also the NIHAO simulations \citep{Wang2015, Buck2020,Buck2020c} have implemented \texttt{CHEMPY} supported yield tables including the TNG yield set \citep{Buck2021} and hence would make for a nice dataset for our analysis. However, we have decided to use the exact same galaxy as in \citet{Philcox_2019} for better comparison of our results.

In detail, we use a single galaxy from the $z = 0$ snapshot of the highest-resolution TNG100-1 simulation. We choose a subhalo (index 523071) with mass close to $10^{12}\,\mathrm{M}_\odot$ to select a Milky Way-like galaxy. From this, we extract a set of 1,000 random `stellar particles' from a total of $\sim 40,000$. Each star particle has a mass of $\sim 1.4\times 10^6 \mathrm{M}_\odot$ \citep{2019ComAC...6....2N}. These act as proxies for stellar environments, giving the elemental mass fractions, $\{d_i^j\}$, and cosmological scale factor, $a_i$, at the time of stellar birth. Mass fractions are converted to [X/Fe] abundance ratios using \citet{2009ARA&A..47..481A} solar abundances as in \texttt{CHEMPY}, with the scale-factor ($a_i$) to birth-time ($T_i$) conversion performed using \texttt{astropy} \citep{astropy:2013,astropy:2018},\footnote{\href{http://www.astropy.org}{http://www.astropy.org}} assuming a $\Lambda$CDM cosmology with \citet{planck2015} parameters, as in TNG \citep{2018MNRAS.475..648P}.\footnote{Note, as for the \texttt{CHEMPY} mock data, we exclude any particles with $T_i\notin[2,12.8]\,\mathrm{Gyr}$ to ensure that the true times are well separated from our training age limits, avoiding neural network errors. This removes $\sim5\%$ of the stars.} Observational errors are incorporated as above, giving a full data-set that is identical in structure to the \texttt{CHEMPY} mock data. For more details and a plot of the [Mg/Fe] vs. [Fe/H] plane for this galaxy see Sec.~6.3 and Fig.~5 of \citet{Philcox_2019}.

We note that this TNG galaxy was deliberately chosen by \citet{Philcox_2019} to have both a high-$\alpha$ and low-$\alpha$ chemical evolution sequence to test their inference on a mock galaxy with Milky Way-like properties. There is still some debate on the exact formation of this bimodality but it is generally attributed to gas-rich mergers and different modes of star formation \citep[e.g.][]{2018MNRAS.474.3629G,2018MNRAS.477.5072M,2019MNRAS.484.3476C,Buck2020,Buck2023}. Similarly, in chemo-dynamical models, Milky Way-like bimodalities can also be achieved by a combination of radial migration and selection effects without the need for mergers or starbursts \citep[e.g.][]{2009MNRAS.396..203S,2013A&A...558A...9M,2017ApJ...835..224A}.

We show our inference results for the TNG data set in Fig.~\ref{fig:TNG_N_star_analysis} and Fig.~\ref{fig:CHEMPY_TNG_sim_sbi}, respectively.
Again, we find that inference becomes better with increasing number of stars. Despite the drastic difference in chemical enrichment model between training and testing data, our SBI pipeline is impressively capable of inferring the correct posterior values. SBI results for  $\log_{10}(N_{\rm Ia})$ are almost perfect while $\alpha_{\rm IMF}$ is slightly biased high. This is very similar to the results of HMC where inference for $\alpha_{\rm IMF}$ is biased and results for $\log_{10}(N_{\rm Ia})$ are more in agreement. Hence, our SBI inference is on par with the HMC results and given their extreme computational advantage they actually supersede HMC.

Looking at the joint posteriors in Figs.~\ref{fig:CHEMPY_TNG_sbi}, \ref{fig:CHEMPY_alt_sbi} and \ref{fig:CHEMPY_TNG_sim_sbi} we see that also in this case SBI recovers the true parameters only with a few $\sigma$ bias similar to HMC. Again, part of this problem is our assumpotion of a factorized posterior that underestimated model uncertainties as for the alternative yield case.

\section{Discussion}
\label{sec: discussion}


Our study demonstrates that simulation-based inference (SBI) provides a powerful and efficient alternative to conventional methods such as Hamiltonian Monte Carlo (HMC) for inferring global galactic parameters. By leveraging neural density estimators and neural network emulators, we achieved remarkable computational efficiency without compromising precision or accuracy.

The key assumption underpinning our methodology -- that individual stars are independently sampled from their respective stellar environments -- is foundational for tractability. However, this assumption warrants closer examination. In reality, stellar abundances are correlated due to shared star formation histories, cumulative enrichment, and dynamical interactions. Future efforts should explore methodologies capable of incorporating such correlations to further refine the accuracy of SBI, potentially through hierarchical modeling or graph-based methods. In particular this assumption leads to over-confident models in the limit of large observational samples that drastically underestimate model uncertainties. One very promising avenue to circumvent this probelm will be compositional score models \citep{compositional_score} which we will explore in future work.

An exciting avenue for future research involves expanding the SBI method presented here to be able to cope with missing data points by switching to a transformer model architecture.
Also, with slight modifications, SBI could be used 
to directly infer empirical nucleosynthetic yields, which remain a major uncertainty in GCE models. In the current framework, \texttt{CHEMPY} relies on tabulated yields from theoretical studies, which may not fully represent the complex processes driving stellar enrichment. Adapting SBI to simultaneously infer galactic parameters and refine empirical yield tables would require changes to the simulator. Specifically, \texttt{CHEMPY} would need to incorporate parameterized yield modifications as part of its input space, allowing for flexible adjustments to enrichment rates during inference. This would also necessitate larger training datasets and enhanced validation techniques to ensure convergence. Such an approach could provide a unified framework for calibrating galactic models directly against observational data. Research in this direction is currently performed and will be part of a future paper.

In the meantime, SBI can also be used to perform model comparison \citep[e.g.][]{model_comp,zhou2024} which can be used to determine which of the tabulated yield sets best match observational constraints. 

Compared to previous HMC-based studies \citep[e.g.][]{Philcox_2019}, our results highlight SBI's resilience to certain types of model misspecification, such as mismatched yield tables. The robustness of SBI in these cases stems from its ability to approximate complex posterior distributions efficiently. Notably, SBI retained high accuracy in scenarios where HMC struggled, particularly for the high-mass slope of the IMF ($\alpha_{\rm ISM}$). This suggests SBI's potential for real-world applications, where the underlying models may deviate from observational data.

The applicability of our method extends to surveys with more observed elements especially neutron capture elements such as GALAH DR4 \citep{Buder2024} and future spectroscopic surveys, such as those planned by the 4MOST \citep{4most} or WEAVE \citep{weave} consortia, which will provide orders of magnitude more data than current datasets. Our findings indicate that SBI can seamlessly scale to such large datasets, offering significant advantages in terms of speed and computational cost.

As anticipated in Section \ref{sec:methods}, the speed of our NN emulator would allow us to perform SNPE, giving up the amortized property in favor of a stronger constraining power, since the Sequential techniques have been shown empirically to outperform the respective amortized version \citep{Ho2024}. This method trains the NDE on a fraction of the initial simulations budget, retrieving a first estimate of the posterior, and at inference time simulate new observations to train on in regions of high posterior density, obtained from the first estimate. In this way we can obtain posteriors that are deliberately optimized for a singular data point. We have decided to leave this approach for future work because the results are quite promising and the amortized property can be crucial for the scalability.

Nevertheless, limitations remain. The simplified physics of the \texttt{CHEMPY} model, while advantageous for computational efficiency, omits the complexities of dynamical gas flows, feedback, and metal mixing present in cosmological simulations. While our tests on IllustrisTNG data affirm the robustness of SBI, integrating more sophisticated models into the inference pipeline represents an exciting avenue for future work. On this line of reasoning we refer also to the discussion in \citet{Philcox_2019}.

\section{Summary and conclusions}
\label{sec: conclusion}

This study introduces simulation-based inference (SBI) as an innovative framework for constraining galactic parameters using stellar chemical abundances. By training neural posterior estimators on forward simulations from the \texttt{CHEMPY} model, we achieved precise and accurate inferences for two key parameters: the high-mass slope of the initial mass function ($\alpha_{\rm IMF}$) and the normalization of Type Ia supernova rates ($\log_{10}(N_{\rm Ia})$).

Our results underscore the transformative advantages of SBI over traditional methods like Hamiltonian Monte Carlo (HMC), marking a paradigm shift in galactic parameter inference:
\begin{itemize}
    \item Orders-of-magnitude speed-up: SBI dramatically reduces computational requirements, achieving runtime improvements exceeding 75,000-fold compared to HMC. For instance, while HMC requires $\sim 42\,\mathrm{hours}$ to infer parameters from just 200 stars, SBI completes inference on thousands of stars in mere minutes. This efficiency makes SBI uniquely suited for analyzing the massive datasets expected from next-generation spectroscopic surveys.
    \item Scalability: SBI's amortized nature allows it to scale seamlessly with the size of the dataset. By training a neural posterior estimator once, the method can be applied repeatedly at virtually no additional computational cost. This scalability is essential for leveraging the millions of stars that future surveys like 4MOST \citep{4most} and WEAVE \citep{weave} will provide, enabling precise population-level inferences.
    \item Robustness to model misspecifications: Unlike HMC, which shows significant biases when faced with discrepancies between model assumptions and data, SBI demonstrates remarkable robustness. Even under conditions of mismatched yield tables or data generated from hydrodynamical simulations, SBI provides accurate and reliable results. This robustness ensures SBI's applicability in real-world scenarios where exact model fidelity cannot be guaranteed.
\end{itemize}

In addition to its immediate advantages, SBI lays the foundation for future advancements in galactic modeling. Its flexibility can enable the direct inference of empirical nucleosynthetic yields and facilitate integration with more complex galaxy formation models. These enhancements will further solidify SBI as a cornerstone method in galactic archaeology.

We note though, that the SBI pipeline presented here is not perfect and suffers from simplified assumptions that have been made in order to make the problem tracktable. In particular, the assumption that the  posterior factorizes is a strong assumption that leads to over-confidence of the models at large observational samples. This is certainly something that needs to be improved in future iterations and methodological work in this direction is currently undergoing. 

In conclusion, SBI represents a breakthrough in simulation-based analysis, delivering unparalleled speed, precision, and scalability. By overcoming the computational limitations of traditional techniques like HMC, SBI paves the way for extracting deeper insights into the chemical and dynamical evolution of galaxies in the era of massive spectroscopic surveys.

\begin{acknowledgements}
      This project was made possible by funding from the Carl Zeiss Stiftung.
\end{acknowledgements}

%
%

\bibliographystyle{aa}
\bibliography{bib.bib}

\begin{thebibliography}{82}
\expandafter\ifx\csname natexlab\endcsname\relax\def\natexlab#1{#1}\fi

\bibitem[{{Abdurro'uf} {et~al.}(2022){Abdurro'uf}, {Accetta}, {Aerts}, {Silva
  Aguirre}, {Ahumada}, {Ajgaonkar}, {Filiz Ak}, {Alam}, {Allende Prieto},
  {Almeida}, {Anders}, {Anderson}, {Andrews}, {Anguiano}, {Aquino-Ort{\'\i}z},
  {Arag{\'o}n-Salamanca}, {Argudo-Fern{\'a}ndez}, {Ata}, {Aubert},
  {Avila-Reese}, {Badenes}, {Barb{\'a}}, {Barger}, {Barrera-Ballesteros},
  {Beaton}, {Beers}, {Belfiore}, {Bender}, {Bernardi}, {Bershady}, {Beutler},
  {Bidin}, {Bird}, {Bizyaev}, {Blanc}, {Blanton}, {Boardman}, {Bolton},
  {Boquien}, {Borissova}, {Bovy}, {Brandt}, {Brown}, {Brownstein}, {Brusa},
  {Buchner}, {Bundy}, {Burchett}, {Bureau}, {Burgasser}, {Cabang}, {Campbell},
  {Cappellari}, {Carlberg}, {Wanderley}, {Carrera}, {Cash}, {Chen}, {Chen},
  {Cherinka}, {Chiappini}, {Choi}, {Chojnowski}, {Chung}, {Clerc}, {Cohen},
  {Comerford}, {Comparat}, {da Costa}, {Covey}, {Crane}, {Cruz-Gonzalez},
  {Culhane}, {Cunha}, {Dai}, {Damke}, {Darling}, {Davidson}, {Davies},
  {Dawson}, {De Lee}, {Diamond-Stanic}, {Cano-D{\'\i}az}, {S{\'a}nchez},
  {Donor}, {Duckworth}, {Dwelly}, {Eisenstein}, {Elsworth}, {Emsellem},
  {Eracleous}, {Escoffier}, {Fan}, {Farr}, {Feng}, {Fern{\'a}ndez-Trincado},
  {Feuillet}, {Filipp}, {Fillingham}, {Frinchaboy}, {Fromenteau}, {Galbany},
  {Garc{\'\i}a}, {Garc{\'\i}a-Hern{\'a}ndez}, {Ge}, {Geisler}, {Gelfand},
  {G{\'e}ron}, {Gibson}, {Goddy}, {Godoy-Rivera}, {Grabowski}, {Green},
  {Greener}, {Grier}, {Griffith}, {Guo}, {Guy}, {Hadjara}, {Harding},
  {Hasselquist}, {Hayes}, {Hearty}, {Hern{\'a}ndez}, {Hill}, {Hogg},
  {Holtzman}, {Horta}, {Hsieh}, {Hsu}, {Hsu}, {Huber}, {Huertas-Company},
  {Hutchinson}, {Hwang}, {Ibarra-Medel}, {Chitham}, {Ilha}, {Imig}, {Jaekle},
  {Jayasinghe}, {Ji}, {Johnson}, {Jones}, {J{\"o}nsson}, {Katkov}, {Khalatyan},
  {Kinemuchi}, {Kisku}, {Knapen}, {Kneib}, {Kollmeier}, {Kong}, {Kounkel},
  {Kreckel}, {Krishnarao}, {Lacerna}, {Lane}, {Langgin}, {Lavender}, {Law},
  {Lazarz}, {Leung}, {Leung}, {Lewis}, {Li}, {Li}, {Lian}, {Liang}, {Lin},
  {Lin}, {Lin}, {Lintott}, {Long}, {Longa-Pe{\~n}a}, {L{\'o}pez-Cob{\'a}},
  {Lu}, {Lundgren}, {Luo}, {Mackereth}, {de la Macorra}, {Mahadevan},
  {Majewski}, {Manchado}, {Mandeville}, {Maraston}, {Margalef-Bentabol},
  {Masseron}, {Masters}, {Mathur}, {McDermid}, {Mckay}, {Merloni},
  {Merrifield}, {Meszaros}, {Miglio}, {Di Mille}, {Minniti}, {Minsley}, \&
  {Monachesi}}]{apogee17}
{Abdurro'uf}, {Accetta}, K., {Aerts}, C., {et~al.} 2022, \apjs, 259, 35

\bibitem[{{Agertz} {et~al.}(2021){Agertz}, {Renaud}, {Feltzing}, {Read},
  {Ryde}, {Andersson}, {Rey}, {Bensby}, \& {Feuillet}}]{Agertz2021}
{Agertz}, O., {Renaud}, F., {Feltzing}, S., {et~al.} 2021, \mnras, 503, 5826

\bibitem[{{Andrews} {et~al.}(2017){Andrews}, {Weinberg}, {Sch{\"o}nrich}, \&
  {Johnson}}]{2017ApJ...835..224A}
{Andrews}, B.~H., {Weinberg}, D.~H., {Sch{\"o}nrich}, R., \& {Johnson}, J.~A.
  2017, \apj, 835, 224

\bibitem[{{Asplund} {et~al.}(2009){Asplund}, {Grevesse}, {Sauval}, \&
  {Scott}}]{2009ARA&A..47..481A}
{Asplund}, M., {Grevesse}, N., {Sauval}, A.~J., \& {Scott}, P. 2009, \araa, 47,
  481

\bibitem[{{Astropy Collaboration} {et~al.}(2013){Astropy Collaboration},
  {Robitaille}, {Tollerud}, {Greenfield}, {Droettboom}, {Bray}, {Aldcroft},
  {Davis}, {Ginsburg}, {Price-Whelan}, {Kerzendorf}, {Conley}, {Crighton},
  {Barbary}, {Muna}, {Ferguson}, {Grollier}, {Parikh}, {Nair}, {Unther},
  {Deil}, {Woillez}, {Conseil}, {Kramer}, {Turner}, {Singer}, {Fox}, {Weaver},
  {Zabalza}, {Edwards}, {Azalee Bostroem}, {Burke}, {Casey}, {Crawford},
  {Dencheva}, {Ely}, {Jenness}, {Labrie}, {Lim}, {Pierfederici}, {Pontzen},
  {Ptak}, {Refsdal}, {Servillat}, \& {Streicher}}]{astropy:2013}
{Astropy Collaboration}, {Robitaille}, T.~P., {Tollerud}, E.~J., {et~al.} 2013,
  aap, 558, A33

\bibitem[{{Bigiel} {et~al.}(2008){Bigiel}, {Leroy}, {Walter}, {Brinks}, {de
  Blok}, {Madore}, \& {Thornley}}]{2008AJ....136.2846B}
{Bigiel}, F., {Leroy}, A., {Walter}, F., {et~al.} 2008, \aj, 136, 2846

\bibitem[{{Buck}(2020)}]{Buck2020}
{Buck}, T. 2020, \mnras, 491, 5435

\bibitem[{{Buck} {et~al.}(2020){Buck}, {Obreja}, {Macci{\`o}}, {Minchev},
  {Dutton}, \& {Ostriker}}]{Buck2020c}
{Buck}, T., {Obreja}, A., {Macci{\`o}}, A.~V., {et~al.} 2020, \mnras, 491, 3461

\bibitem[{{Buck} {et~al.}(2023){Buck}, {Obreja}, {Ratcliffe}, {Lu}, {Minchev},
  \& {Macci{\`o}}}]{Buck2023}
{Buck}, T., {Obreja}, A., {Ratcliffe}, B., {et~al.} 2023, \mnras, 523, 1565

\bibitem[{{Buck} {et~al.}(2021){Buck}, {Rybizki}, {Buder}, {Obreja},
  {Macci{\`o}}, {Pfrommer}, {Steinmetz}, \& {Ness}}]{Buck2021}
{Buck}, T., {Rybizki}, J., {Buder}, S., {et~al.} 2021, \mnras, 508, 3365

\bibitem[{{Buder} {et~al.}(2024){Buder}, {Kos}, {Wang}, {McKenzie}, {Howell},
  {Martell}, {Hayden}, {Zucker}, {Nordlander}, {Montet}, {Traven},
  {Bland-Hawthorn}, {De Silva}, {Freeman}, {Lewis}, {Lind}, {Sharma},
  {Simpson}, {Stello}, {Zwitter}, {Amarsi}, {Armstrong}, {Banks}, {Beavis},
  {Beeson}, {Chen}, {Ciuc{\u{a}}}, {Da Costa}, {de Grijs}, {Martin}, {Nataf},
  {Ness}, {Rains}, {Scarr}, {Vogrin{\v{c}}i{\v{c}}}, {Wang}, {Wittenmyer},
  {Xie}, \& {The GALAH Collaboration}}]{Buder2024}
{Buder}, S., {Kos}, J., {Wang}, E.~X., {et~al.} 2024, arXiv e-prints,
  arXiv:2409.19858

\bibitem[{{Buder} {et~al.}(2021){Buder}, {Sharma}, {Kos}, {Amarsi},
  {Nordlander}, {Lind}, {Martell}, {Asplund}, {Bland-Hawthorn}, {Casey}, {de
  Silva}, {D'Orazi}, {Freeman}, {Hayden}, {Lewis}, {Lin}, {Schlesinger},
  {Simpson}, {Stello}, {Zucker}, {Zwitter}, {Beeson}, {Buck}, {Casagrande},
  {Clark}, {{\v{C}}otar}, {da Costa}, {de Grijs}, {Feuillet}, {Horner},
  {Kafle}, {Khanna}, {Kobayashi}, {Liu}, {Montet}, {Nandakumar}, {Nataf},
  {Ness}, {Spina}, {Tepper-Garc{\'\i}a}, {Ting}, {Traven},
  {Vogrin{\v{c}}i{\v{c}}}, {Wittenmyer}, {Wyse}, {{\v{Z}}erjal},
  {{\v{Z}}erjal}, \& {Galah Collaboration}}]{Buder2021}
{Buder}, S., {Sharma}, S., {Kos}, J., {et~al.} 2021, \mnras, 506, 150

\bibitem[{{Chabrier}(2003)}]{2003PASP..115..763C}
{Chabrier}, G. 2003, \pasp, 115, 763

\bibitem[{{Chabrier} {et~al.}(2014){Chabrier}, {Hennebelle}, \&
  {Charlot}}]{Chabrier2014}
{Chabrier}, G., {Hennebelle}, P., \& {Charlot}, S. 2014, \apj, 796, 75

\bibitem[{{Clarke} {et~al.}(2019){Clarke}, {Debattista}, {Nidever}, {Loebman},
  {Simons}, {Kassin}, {Du}, {Ness}, {Fisher}, {Quinn}, {Wadsley}, {Freeman}, \&
  {Popescu}}]{2019MNRAS.484.3476C}
{Clarke}, A.~J., {Debattista}, V.~P., {Nidever}, D.~L., {et~al.} 2019, \mnras,
  484, 3476

\bibitem[{{Clauwens} {et~al.}(2016){Clauwens}, {Schaye}, \&
  {Franx}}]{2016MNRAS.462.2832C}
{Clauwens}, B., {Schaye}, J., \& {Franx}, M. 2016, \mnras, 462, 2832

\bibitem[{{Cook} {et~al.}(2006){Cook}, {Gelman}, \& {Rubin}}]{Cook2006}
{Cook}, S.~R., {Gelman}, A., \& {Rubin}, D.~B. 2006, Journal of Computational
  and Graphical Statistics, 15, 675

\bibitem[{{C{\^o}t{\'e}} \& {Ritter}(2018)}]{omega2018}
{C{\^o}t{\'e}}, B. \& {Ritter}, C. 2018, {OMEGA: One-zone Model for the
  Evolution of GAlaxies}, Astrophysics Source Code Library, record
  ascl:1806.018

\bibitem[{{C{\^o}t{\'e}} {et~al.}(2016){C{\^o}t{\'e}}, {Ritter}, {O'Shea},
  {Herwig}, {Pignatari}, {Jones}, \& {Fryer}}]{2016ApJ...824...82C}
{C{\^o}t{\'e}}, B., {Ritter}, C., {O'Shea}, B.~W., {et~al.} 2016, \apj, 824, 82

\bibitem[{Cranmer {et~al.}(2020)Cranmer, Brehmer, \& Louppe}]{Cranmer2020}
Cranmer, K., Brehmer, J., \& Louppe, G. 2020, Proceedings of the National
  Academy of Sciences, 117, 30055

\bibitem[{{Dalton} {et~al.}(2018){Dalton}, {Trager}, {Abrams}, {Bonifacio},
  {Aguerri}, {Vallenari}, {Middleton}, {Benn}, {Dee}, {Say{\`e}de}, {Lewis},
  {Pragt}, {Pic{\'o}}, {Walton}, {Rey}, {Allende Prieto}, {Lhom{\'e}},
  {Terrett}, {Brock}, {Gilbert}, {Ridings}, {Verheijen}, {Tosh}, {Steele},
  {Stuik}, {Kroes}, {Tromp}, {Kragt}, {Lesman}, {Mottram}, {Bates}, {Gribbin},
  {Burgal}, {Herreros}, {Delgado}, {Martin}, {Cano}, {Navarro}, {Irwin},
  {Lewis}, {Gonzales Solares}, {O'Mahony}, {Bianco}, {Zurita}, {ter Horst},
  {Molinari}, {Lodi}, {Guerra}, {Baruffolo}, {Carrasco}, {Farkas}, {Schallig},
  {Hill}, {Smith}, {Drew}, {Poggianti}, {Pieri}, {Jin}, {Dominquez Palmero},
  {Fari{\~n}a}, {Martin}, {Worley}, {Murphy}, {Hidalgo}, {Mignot}, {Bishop},
  {Guest}, {Elswijk}, {de Haan}, {Hanenburg}, {Salasnich}, {Mayya},
  {Izazaga-P{\'e}rez}, \& {Peralta de Arriba}}]{weave}
{Dalton}, G., {Trager}, S., {Abrams}, D.~C., {et~al.} 2018, in Society of
  Photo-Optical Instrumentation Engineers (SPIE) Conference Series, Vol. 10702,
  Ground-based and Airborne Instrumentation for Astronomy VII, ed. C.~J.
  {Evans}, L.~{Simard}, \& H.~{Takami}, 107021B

\bibitem[{{de Jong} {et~al.}(2014){de Jong}, {Barden}, {Bellido-Tirado},
  {Brynnel}, {Chiappini}, {Depagne}, {Haynes}, {Johl}, {Phillips}, {Schnurr},
  {Schwope}, {Walcher}, {Bauer}, {Cescutti}, {Cioni}, {Dionies}, {Enke},
  {Haynes}, {Kelz}, {Kitaura}, {Lamer}, {Minchev}, {M{\"u}ller}, {Nuza},
  {Olaya}, {Piffl}, {Popow}, {Saviauk}, {Steinmetz}, {Ural}, {Valentini},
  {Winkler}, {Wisotzki}, {Ansorge}, {Banerji}, {Gonzalez Solares}, {Irwin},
  {Kennicutt}, {King}, {McMahon}, {Koposov}, {Parry}, {Sun}, {Walton},
  {Finger}, {Iwert}, {Krumpe}, {Lizon}, {Mainieri}, {Amans}, {Bonifacio},
  {Cohen}, {Fran{\c{c}}ois}, {Jagourel}, {Mignot}, {Royer}, {Sartoretti},
  {Bender}, {Hess}, {Lang-Bardl}, {Muschielok}, {Schlichter}, {B{\"o}hringer},
  {Boller}, {Bongiorno}, {Brusa}, {Dwelly}, {Merloni}, {Nandra}, {Salvato},
  {Pragt}, {Navarro}, {Gerlofsma}, {Roelfsema}, {Dalton}, {Middleton}, {Tosh},
  {Boeche}, {Caffau}, {Christlieb}, {Grebel}, {Hansen}, {Koch}, {Ludwig},
  {Mandel}, {Quirrenbach}, {Sbordone}, {Seifert}, {Thimm}, {Helmi}, {trager},
  {Bensby}, {Feltzing}, {Ruchti}, {Edvardsson}, {Korn}, {Lind}, {Boland},
  {Colless}, {Frost}, {Gilbert}, {Gillingham}, {Lawrence}, {Legg}, {Saunders},
  {Sheinis}, {Driver}, {Robotham}, {Bacon}, {Caillier}, {Kosmalski}, {Laurent},
  \& {Richard}}]{4most}
{de Jong}, R.~S., {Barden}, S., {Bellido-Tirado}, O., {et~al.} 2014, in Society
  of Photo-Optical Instrumentation Engineers (SPIE) Conference Series, Vol.
  9147, Ground-based and Airborne Instrumentation for Astronomy V, ed. S.~K.
  {Ramsay}, I.~S. {McLean}, \& H.~{Takami}, 91470M

\bibitem[{{Defazio} {et~al.}(2024){Defazio}, {Yang}, {Mehta}, {Mishchenko},
  {Khaled}, \& {Cutkosky}}]{schedulefree}
{Defazio}, A., {Yang}, X.~A., {Mehta}, H., {et~al.} 2024, arXiv e-prints,
  arXiv:2405.15682

\bibitem[{{Doherty} {et~al.}(2014){Doherty}, {Gil-Pons}, {Lau}, {Lattanzio}, \&
  {Siess}}]{2014MNRAS.437..195D}
{Doherty}, C.~L., {Gil-Pons}, P., {Lau}, H. H.~B., {Lattanzio}, J.~C., \&
  {Siess}, L. 2014, \mnras, 437, 195

\bibitem[{Durkan {et~al.}(2019)Durkan, Bekasov, Murray, \&
  Papamakarios}]{durkan2019neuralsplineflows}
Durkan, C., Bekasov, A., Murray, I., \& Papamakarios, G. 2019, Neural Spline
  Flows

\bibitem[{{Fishlock} {et~al.}(2014){Fishlock}, {Karakas}, {Lugaro}, \&
  {Yong}}]{2014ApJ...797...44F}
{Fishlock}, C.~K., {Karakas}, A.~I., {Lugaro}, M., \& {Yong}, D. 2014, \apj,
  797, 44

\bibitem[{{Font} {et~al.}(2020){Font}, {McCarthy}, {Poole-Mckenzie},
  {Stafford}, {Brown}, {Schaye}, {Crain}, {Theuns}, \& {Schaller}}]{Font2020}
{Font}, A.~S., {McCarthy}, I.~G., {Poole-Mckenzie}, R., {et~al.} 2020, \mnras,
  498, 1765

\bibitem[{{Geffner} {et~al.}(2022){Geffner}, {Papamakarios}, \&
  {Mnih}}]{compositional_score}
{Geffner}, T., {Papamakarios}, G., \& {Mnih}, A. 2022, arXiv e-prints,
  arXiv:2209.14249

\bibitem[{Gloeckler {et~al.}(2024)Gloeckler, Deistler, Weilbach, Wood, \&
  Macke}]{Gloeckler2024AllinoneSI}
Gloeckler, M., Deistler, M., Weilbach, C., Wood, F., \& Macke, J.~H. 2024,
  ArXiv, abs/2404.09636

\bibitem[{{Grand} {et~al.}(2018){Grand}, {Bustamante}, {G{\'o}mez}, {Kawata},
  {Marinacci}, {Pakmor}, {Rix}, {Simpson}, {Sparre}, \&
  {Springel}}]{2018MNRAS.474.3629G}
{Grand}, R.~J.~J., {Bustamante}, S., {G{\'o}mez}, F.~A., {et~al.} 2018, \mnras,
  474, 3629

\bibitem[{{Griffith} {et~al.}(2019){Griffith}, {Johnson}, \&
  {Weinberg}}]{2019arXiv190806113G}
{Griffith}, E., {Johnson}, J.~A., \& {Weinberg}, D.~H. 2019, arXiv e-prints,
  arXiv:1908.06113

\bibitem[{{Gutcke} \& {Springel}(2019)}]{2019MNRAS.482..118G}
{Gutcke}, T.~A. \& {Springel}, V. 2019, \mnras, 482, 118

\bibitem[{{Ho} {et~al.}(2024){Ho}, {Bartlett}, {Chartier}, {Cuesta-Lazaro},
  {Ding}, {Lapel}, {Lemos}, {Lovell}, {Makinen}, {Modi}, {Pandya}, {Pandey},
  {Perez}, {Wandelt}, \& {Bryan}}]{Ho2024}
{Ho}, M., {Bartlett}, D.~J., {Chartier}, N., {et~al.} 2024, The Open Journal of
  Astrophysics, 7, 54

\bibitem[{{Hopkins} {et~al.}(2018){Hopkins}, {Wetzel}, {Kere{\v{s}}},
  {Faucher-Gigu{\`e}re}, {Quataert}, {Boylan-Kolchin}, {Murray}, {Hayward},
  {Garrison-Kimmel}, {Hummels}, {Feldmann}, {Torrey}, {Ma},
  {Angl{\'e}s-Alc{\'a}zar}, {Su}, {Orr}, {Schmitz}, {Escala}, {Sanderson},
  {Grudi{\'c}}, {Hafen}, {Kim}, {Fitts}, {Bullock}, {Wheeler}, {Chan},
  {Elbert}, \& {Narayanan}}]{Hopkins2018}
{Hopkins}, P.~F., {Wetzel}, A., {Kere{\v{s}}}, D., {et~al.} 2018, \mnras, 480,
  800

\bibitem[{{Jim{\'e}nez} {et~al.}(2015){Jim{\'e}nez}, {Tissera}, \&
  {Matteucci}}]{2015ApJ...810..137J}
{Jim{\'e}nez}, N., {Tissera}, P.~B., \& {Matteucci}, F. 2015, \apj, 810, 137

\bibitem[{{Johnson} \& {Weinberg}(2020)}]{vice}
{Johnson}, J.~W. \& {Weinberg}, D.~H. 2020, \mnras, 498, 1364

\bibitem[{{Karakas}(2010)}]{2010MNRAS.403.1413K}
{Karakas}, A.~I. 2010, \mnras, 403, 1413

\bibitem[{{Karakas} \& {Lugaro}(2016)}]{2016ApJ...825...26K}
{Karakas}, A.~I. \& {Lugaro}, M. 2016, \apj, 825, 26

\bibitem[{{Kobayashi} {et~al.}(2006){Kobayashi}, {Umeda}, {Nomoto}, {Tominaga},
  \& {Ohkubo}}]{2006ApJ...653.1145K}
{Kobayashi}, C., {Umeda}, H., {Nomoto}, K., {Tominaga}, N., \& {Ohkubo}, T.
  2006, \apj, 653, 1145

\bibitem[{{Lemos} {et~al.}(2023){Lemos}, {Coogan}, {Hezaveh}, \&
  {Perreault-Levasseur}}]{Lemos2023}
{Lemos}, P., {Coogan}, A., {Hezaveh}, Y., \& {Perreault-Levasseur}, L. 2023,
  40th International Conference on Machine Learning, 202, 19256

\bibitem[{{Luger} {et~al.}(2021){Luger}, {Bedell}, {Foreman-Mackey},
  {Crossfield}, {Zhao}, \& {Hogg}}]{Luger2021}
{Luger}, R., {Bedell}, M., {Foreman-Mackey}, D., {et~al.} 2021, arXiv e-prints,
  arXiv:2110.06271

\bibitem[{{Mackereth} {et~al.}(2018){Mackereth}, {Crain}, {Schiavon}, {Schaye},
  {Theuns}, \& {Schaller}}]{2018MNRAS.477.5072M}
{Mackereth}, J.~T., {Crain}, R.~A., {Schiavon}, R.~P., {et~al.} 2018, \mnras,
  477, 5072

\bibitem[{{Majewski} {et~al.}(2016){Majewski}, {APOGEE Team}, \& {APOGEE-2
  Team}}]{Majewski2016}
{Majewski}, S.~R., {APOGEE Team}, \& {APOGEE-2 Team}. 2016, Astronomische
  Nachrichten, 337, 863

\bibitem[{{Maoz} \& {Mannucci}(2012)}]{2012PASA...29..447M}
{Maoz}, D. \& {Mannucci}, F. 2012, \pasa, 29, 447

\bibitem[{{Maoz} {et~al.}(2012){Maoz}, {Mannucci}, \&
  {Brandt}}]{2012MNRAS.426.3282M}
{Maoz}, D., {Mannucci}, F., \& {Brandt}, T.~D. 2012, \mnras, 426, 3282

\bibitem[{{Maoz} {et~al.}(2010){Maoz}, {Sharon}, \&
  {Gal-Yam}}]{2010ApJ...722.1879M}
{Maoz}, D., {Sharon}, K., \& {Gal-Yam}, A. 2010, \apj, 722, 1879

\bibitem[{{Mart{\'\i}n-Navarro} {et~al.}(2019){Mart{\'\i}n-Navarro},
  {Lyubenova}, {van de Ven}, {Falc{\'o}n-Barroso}, {Coccato}, {Corsini},
  {Gadotti}, {Iodice}, {La Barbera}, {McDermid}, {Pinna}, {Sarzi}, {Viaene},
  {de Zeeuw}, \& {Zhu}}]{Martin2019}
{Mart{\'\i}n-Navarro}, I., {Lyubenova}, M., {van de Ven}, G., {et~al.} 2019,
  \aap, 626, A124

\bibitem[{{Minchev} {et~al.}(2013){Minchev}, {Chiappini}, \&
  {Martig}}]{2013A&A...558A...9M}
{Minchev}, I., {Chiappini}, C., \& {Martig}, M. 2013, \aap, 558, A9

\bibitem[{{Moll{\'a}} {et~al.}(2015){Moll{\'a}}, {Cavichia}, {Gavil{\'a}n}, \&
  {Gibson}}]{2015MNRAS.451.3693M}
{Moll{\'a}}, M., {Cavichia}, O., {Gavil{\'a}n}, M., \& {Gibson}, B.~K. 2015,
  \mnras, 451, 3693

\bibitem[{{Nelson} {et~al.}(2019){Nelson}, {Springel}, {Pillepich},
  {Rodriguez-Gomez}, {Torrey}, {Genel}, {Vogelsberger}, {Pakmor}, {Marinacci},
  {Weinberger}, {Kelley}, {Lovell}, {Diemer}, \&
  {Hernquist}}]{2019ComAC...6....2N}
{Nelson}, D., {Springel}, V., {Pillepich}, A., {et~al.} 2019, Computational
  Astrophysics and Cosmology, 6, 2

\bibitem[{{Ness} {et~al.}(2019){Ness}, {Johnston}, {Blancato}, {Rix}, {Beane},
  {Bird}, \& {Hawkins}}]{2019arXiv190710606N}
{Ness}, M.~K., {Johnston}, K.~V., {Blancato}, K., {et~al.} 2019, arXiv
  e-prints, arXiv:1907.10606

\bibitem[{{Nomoto} {et~al.}(1997){Nomoto}, {Iwamoto}, {Nakasato}, {Thielemann},
  {Brachwitz}, {Tsujimoto}, {Kubo}, \& {Kishimoto}}]{1997NuPhA.621..467N}
{Nomoto}, K., {Iwamoto}, K., {Nakasato}, N., {et~al.} 1997, \nphysa, 621, 467

\bibitem[{{Nomoto} {et~al.}(2013){Nomoto}, {Kobayashi}, \&
  {Tominaga}}]{Nomoto2013}
{Nomoto}, K., {Kobayashi}, C., \& {Tominaga}, N. 2013, \araa, 51, 457

\bibitem[{Papamakarios {et~al.}(2021)Papamakarios, Nalisnick, Rezende, Mohamed,
  \& Lakshminarayanan}]{Papamakarios:2021}
Papamakarios, G., Nalisnick, E., Rezende, D.~J., Mohamed, S., \&
  Lakshminarayanan, B. 2021, Normalizing Flows for Probabilistic Modeling and
  Inference

\bibitem[{Papamakarios {et~al.}(2018)Papamakarios, Pavlakou, \&
  Murray}]{papamakarios2018maskedautoregressiveflowdensity}
Papamakarios, G., Pavlakou, T., \& Murray, I. 2018, Masked Autoregressive Flow
  for Density Estimation

\bibitem[{Philcox {et~al.}(2018)Philcox, Rybizki, \& Gutcke}]{Philcox_2018}
Philcox, O., Rybizki, J., \& Gutcke, T.~A. 2018, The Astrophysical Journal,
  861, 40

\bibitem[{{Philcox} {et~al.}(2018){Philcox}, {Rybizki}, \&
  {Gutcke}}]{2018ApJ...861...40P}
{Philcox}, O., {Rybizki}, J., \& {Gutcke}, T.~A. 2018, \apj, 861, 40

\bibitem[{Philcox \& Rybizki(2019)}]{Philcox_2019}
Philcox, O. H.~E. \& Rybizki, J. 2019, The Astrophysical Journal, 887, 9

\bibitem[{{Pillepich} {et~al.}(2018{\natexlab{a}}){Pillepich}, {Nelson},
  {Hernquist}, {Springel}, {Pakmor}, {Torrey}, {Weinberger}, {Genel}, {Naiman},
  {Marinacci}, \& {Vogelsberger}}]{2018MNRAS.475..648P}
{Pillepich}, A., {Nelson}, D., {Hernquist}, L., {et~al.} 2018{\natexlab{a}},
  \mnras, 475, 648

\bibitem[{{Pillepich} {et~al.}(2018{\natexlab{b}}){Pillepich}, {Springel},
  {Nelson}, {Genel}, {Naiman}, {Pakmor}, {Hernquist}, {Torrey}, {Vogelsberger},
  {Weinberger}, \& {Marinacci}}]{Pillepich2018}
{Pillepich}, A., {Springel}, V., {Nelson}, D., {et~al.} 2018{\natexlab{b}},
  \mnras, 473, 4077

\bibitem[{{Pillepich} {et~al.}(2018{\natexlab{c}}){Pillepich}, {Springel},
  {Nelson}, {Genel}, {Naiman}, {Pakmor}, {Hernquist}, {Torrey}, {Vogelsberger},
  {Weinberger}, \& {Marinacci}}]{2018MNRAS.473.4077P}
{Pillepich}, A., {Springel}, V., {Nelson}, D., {et~al.} 2018{\natexlab{c}},
  \mnras, 473, 4077

\bibitem[{{Planck Collaboration} {et~al.}(2016){Planck Collaboration}, {Ade},
  {Aghanim}, {Arnaud}, {Ashdown}, {Aumont}, {Baccigalupi}, {Banday},
  {Barreiro}, {Bartlett}, {Bartolo}, {Battaner}, {Battye}, {Benabed},
  {Beno{\^\i}t}, {Benoit-L{\'e}vy}, {Bernard}, {Bersanelli}, {Bielewicz},
  {Bock}, {Bonaldi}, {Bonavera}, {Bond}, {Borrill}, {Bouchet}, {Boulanger},
  {Bucher}, {Burigana}, {Butler}, {Calabrese}, {Cardoso}, {Catalano},
  {Challinor}, {Chamballu}, {Chary}, {Chiang}, {Chluba}, {Christensen},
  {Church}, {Clements}, {Colombi}, {Colombo}, {Combet}, {Coulais}, {Crill},
  {Curto}, {Cuttaia}, {Danese}, {Davies}, {Davis}, {de Bernardis}, {de Rosa},
  {de Zotti}, {Delabrouille}, {D{\'e}sert}, {Di Valentino}, {Dickinson},
  {Diego}, {Dolag}, {Dole}, {Donzelli}, {Dor{\'e}}, {Douspis}, {Ducout},
  {Dunkley}, {Dupac}, {Efstathiou}, {Elsner}, {En{\ss}lin}, {Eriksen},
  {Farhang}, {Fergusson}, {Finelli}, {Forni}, {Frailis}, {Fraisse},
  {Franceschi}, {Frejsel}, {Galeotta}, {Galli}, {Ganga}, {Gauthier}, {Gerbino},
  {Ghosh}, {Giard}, {Giraud-H{\'e}raud}, {Giusarma}, {Gjerl{\o}w},
  {Gonz{\'a}lez-Nuevo}, {G{\'o}rski}, {Gratton}, {Gregorio}, {Gruppuso},
  {Gudmundsson}, {Hamann}, {Hansen}, {Hanson}, {Harrison}, {Helou}, {Henrot-
  Versill{\'e}}, {Hern{\'a}ndez-Monteagudo}, {Herranz}, {Hildebrandt}, {Hivon},
  {Hobson}, {Holmes}, {Hornstrup}, {Hovest}, {Huang}, {Huffenberger}, {Hurier},
  {Jaffe}, {Jaffe}, {Jones}, {Juvela}, {Keih{\"a}nen}, {Keskitalo}, {Kisner},
  {Kneissl}, {Knoche}, {Knox}, {Kunz}, {Kurki-Suonio}, {Lagache},
  {L{\"a}hteenm{\"a}ki}, {Lamarre}, {Lasenby}, {Lattanzi}, {Lawrence}, {Leahy},
  {Leonardi}, {Lesgourgues}, {Levrier}, {Lewis}, {Liguori}, {Lilje},
  {Linden-V{\o}rnle}, {L{\'o}pez-Caniego}, {Lubin}, {Mac{\'\i}as-P{\'e}rez},
  {Maggio}, {Maino}, {Mandolesi}, {Mangilli}, {Marchini}, {Maris}, {Martin},
  {Martinelli}, {Mart{\'\i}nez-Gonz{\'a}lez}, {Masi}, {Matarrese}, {McGehee},
  {Meinhold}, {Melchiorri}, {Melin}, {Mendes}, {Mennella}, {Migliaccio},
  {Millea}, {Mitra}, {Miville-Desch{\^e}nes}, {Moneti}, {Montier}, {Morgante},
  {Mortlock}, {Moss}, {Munshi}, {Murphy}, {Naselsky}, {Nati}, {Natoli},
  {Netterfield}, {N{\o}rgaard-Nielsen}, {Noviello}, {Novikov}, {Novikov},
  {Oxborrow}, {Paci}, {Pagano}, {Pajot}, {Paladini}, {Paoletti}, {Partridge},
  {Pasian}, {Patanchon}, {Pearson}, {Perdereau}, {Perotto}, {Perrotta},
  {Pettorino}, {Piacentini}, {Piat}, {Pierpaoli}, {Pietrobon}, {Plaszczynski},
  {Pointecouteau}, {Polenta}, {Popa}, {Pratt}, {Pr{\'e}zeau}, {Prunet},
  {Puget}, {Rachen}, {Reach}, {Rebolo}, {Reinecke}, {Remazeilles}, {Renault},
  {Renzi}, {Ristorcelli}, {Rocha}, {Rosset}, {Rossetti}, {Roudier},
  {Rouill{\'e} d'Orfeuil}, {Rowan-Robinson}, {Rubi{\~n}o-Mart{\'\i}n},
  {Rusholme}, {Said}, {Salvatelli}, {Salvati}, {Sandri}, {Santos},
  {Savelainen}, {Savini}, {Scott}, {Seiffert}, {Serra}, {Shellard}, {Spencer},
  {Spinelli}, {Stolyarov}, {Stompor}, {Sudiwala}, {Sunyaev}, {Sutton},
  {Suur-Uski}, {Sygnet}, {Tauber}, {Terenzi}, {Toffolatti}, {Tomasi},
  {Tristram}, {Trombetti}, {Tucci}, {Tuovinen}, {T{\"u}rler}, {Umana},
  {Valenziano}, {Valiviita}, {Van Tent}, {Vielva}, {Villa}, {Wade}, {Wandelt},
  {Wehus}, {White}, {White}, {Wilkinson}, {Yvon}, {Zacchei}, \&
  {Zonca}}]{planck2015}
{Planck Collaboration}, {Ade}, P.~A.~R., {Aghanim}, N., {et~al.} 2016, \aap,
  594, A13

\bibitem[{{Portinari} {et~al.}(1998){Portinari}, {Chiosi}, \&
  {Bressan}}]{portinari}
{Portinari}, L., {Chiosi}, C., \& {Bressan}, A. 1998, \aap, 334, 505

\bibitem[{{Price-Whelan} {et~al.}(2018){Price-Whelan}, {Sip{'{o}}cz},
  {G{"u}nther}, {Lim}, {Crawford}, {Conseil}, {Shupe}, {Craig}, {Dencheva},
  {Ginsburg}, {VanderPlas}, {Bradley}, {P{'e}rez-Su{'a}rez}, {de Val-Borro},
  {Paper Contributors}, {Aldcroft}, {Cruz}, {Robitaille}, {Tollerud},
  {Coordination Committee}, {Ardelean}, {Babej}, {Bach}, {Bachetti}, {Bakanov},
  {Bamford}, {Barentsen}, {Barmby}, {Baumbach}, {Berry}, {Biscani}, {Boquien},
  {Bostroem}, {Bouma}, {Brammer}, {Bray}, {Breytenbach}, {Buddelmeijer},
  {Burke}, {Calderone}, {Cano Rodr{'i}guez}, {Cara}, {Cardoso}, {Cheedella},
  {Copin}, {Corrales}, {Crichton}, {D{ extquoteright}Avella}, {Deil},
  {Depagne}, {Dietrich}, {Donath}, {Droettboom}, {Earl}, {Erben}, {Fabbro},
  {Ferreira}, {Finethy}, {Fox}, {Garrison}, {Gibbons}, {Goldstein}, {Gommers},
  {Greco}, {Greenfield}, {Groener}, {Grollier}, {Hagen}, {Hirst}, {Homeier},
  {Horton}, {Hosseinzadeh}, {Hu}, {Hunkeler}, {Ivezi{'c}}, {Jain}, {Jenness},
  {Kanarek}, {Kendrew}, {Kern}, {Kerzendorf}, {Khvalko}, {King}, {Kirkby},
  {Kulkarni}, {Kumar}, {Lee}, {Lenz}, {Littlefair}, {Ma}, {Macleod},
  {Mastropietro}, {McCully}, {Montagnac}, {Morris}, {Mueller}, {Mumford},
  {Muna}, {Murphy}, {Nelson}, {Nguyen}, {Ninan}, {N{"o}the}, {Ogaz}, {Oh},
  {Parejko}, {Parley}, {Pascual}, {Patil}, {Patil}, {Plunkett}, {Prochaska},
  {Rastogi}, {Reddy Janga}, {Sabater}, {Sakurikar}, {Seifert}, {Sherbert},
  {Sherwood-Taylor}, {Shih}, {Sick}, {Silbiger}, {Singanamalla}, {Singer},
  {Sladen}, {Sooley}, {Sornarajah}, {Streicher}, {Teuben}, {Thomas},
  {Tremblay}, {Turner}, {Terr{'o}n}, {van Kerkwijk}, {de la Vega}, {Watkins},
  {Weaver}, {Whitmore}, {Woillez}, {Zabalza}, \& {Contributors}}]{astropy:2018}
{Price-Whelan}, A.~M., {Sip{'{o}}cz}, B.~M., {G{"u}nther}, H.~M., {et~al.}
  2018, aj, 156, 123

\bibitem[{{Romano} {et~al.}(2005){Romano}, {Chiappini}, {Matteucci}, \&
  {Tosi}}]{2005A&A...430..491R}
{Romano}, D., {Chiappini}, C., {Matteucci}, F., \& {Tosi}, M. 2005, \aap, 430,
  491

\bibitem[{{Rybizki} \& {Just}(2015)}]{Rybizki2015}
{Rybizki}, J. \& {Just}, A. 2015, \mnras, 447, 3880

\bibitem[{Rybizki {et~al.}(2017)Rybizki, Just, \& Rix}]{Rybizki_2017}
Rybizki, J., Just, A., \& Rix, H.-W. 2017, \aap, 605, A59

\bibitem[{{Sawala} {et~al.}(2016){Sawala}, {Frenk}, {Fattahi}, {Navarro},
  {Bower}, {Crain}, {Dalla Vecchia}, {Furlong}, {Helly}, {Jenkins}, {Oman},
  {Schaller}, {Schaye}, {Theuns}, {Trayford}, \& {White}}]{Sawala2016}
{Sawala}, T., {Frenk}, C.~S., {Fattahi}, A., {et~al.} 2016, \mnras, 457, 1931

\bibitem[{{Sch{\"o}nrich} \& {Binney}(2009)}]{2009MNRAS.396..203S}
{Sch{\"o}nrich}, R. \& {Binney}, J. 2009, \mnras, 396, 203

\bibitem[{{Spurio Mancini} {et~al.}(2023){Spurio Mancini}, {Docherty}, {Price},
  \& {McEwen}}]{model_comp}
{Spurio Mancini}, A., {Docherty}, M.~M., {Price}, M.~A., \& {McEwen}, J.~D.
  2023, RAS Techniques and Instruments, 2, 710

\bibitem[{{Talbot} \& {Arnett}(1971)}]{Talbot1971}
{Talbot}, Jr., R.~J. \& {Arnett}, W.~D. 1971, \apj, 170, 409

\bibitem[{{Talts} {et~al.}(2018){Talts}, {Betancourt}, {Simpson}, {Vehtari}, \&
  {Gelman}}]{Talts2018}
{Talts}, S., {Betancourt}, M., {Simpson}, D., {Vehtari}, A., \& {Gelman}, A.
  2018, arXiv e-prints, arXiv:1804.06788

\bibitem[{{Thielemann} {et~al.}(2003){Thielemann}, {Argast}, {Brachwitz},
  {Hix}, {H{\"o}flich}, {Liebend{\"o}rfer}, {Martinez-Pinedo}, {Mezzacappa},
  {Panov}, \& {Rauscher}}]{2003NuPhA.718..139T}
{Thielemann}, F.~K., {Argast}, D., {Brachwitz}, F., {et~al.} 2003, \nphysa,
  718, 139

\bibitem[{{van Dokkum} {et~al.}(2013){van Dokkum}, {Leja}, {Nelson}, {Patel},
  {Skelton}, {Momcheva}, {Brammer}, {Whitaker}, {Lundgren}, {Fumagalli},
  {Conroy}, {F{\"o}rster Schreiber}, {Franx}, {Kriek}, {Labb{\'e}},
  {Marchesini}, {Rix}, {van der Wel}, \& {Wuyts}}]{2013ApJ...771L..35V}
{van Dokkum}, P.~G., {Leja}, J., {Nelson}, E.~J., {et~al.} 2013, \apjl, 771,
  L35

\bibitem[{{Vincenzo} {et~al.}(2015){Vincenzo}, {Matteucci}, {Recchi}, {Calura},
  {McWilliam}, \& {Lanfranchi}}]{2015MNRAS.449.1327V}
{Vincenzo}, F., {Matteucci}, F., {Recchi}, S., {et~al.} 2015, \mnras, 449, 1327

\bibitem[{{Viterbo} \& {Buck}(2024)}]{Viterbo2024}
{Viterbo}, G. \& {Buck}, T. 2024, arXiv e-prints, arXiv:2411.17269

\bibitem[{{Wang} {et~al.}(2024){Wang}, {Carrillo}, {Ness}, \&
  {Buck}}]{Wang2024}
{Wang}, K., {Carrillo}, A., {Ness}, M.~K., \& {Buck}, T. 2024, \mnras, 527, 321

\bibitem[{{Wang} {et~al.}(2015){Wang}, {Dutton}, {Stinson}, {Macci{\`o}},
  {Penzo}, {Kang}, {Keller}, \& {Wadsley}}]{Wang2015}
{Wang}, L., {Dutton}, A.~A., {Stinson}, G.~S., {et~al.} 2015, \mnras, 454, 83

\bibitem[{{Weinberg} {et~al.}(2019){Weinberg}, {Holtzman}, {Hasselquist},
  {Bird}, {Johnson}, {Shetrone}, {Sobeck}, {Allende Prieto}, {Bizyaev},
  {Carrera}, {Cohen}, {Cunha}, {Ebelke}, {Fernandez-Trincado},
  {Garc{\'\i}a-Hern{\'a}ndez}, {Hayes}, {J{\"o}nsson}, {Lane}, {Majewski},
  {Malanushenko}, {M{\'e}sz{\'a}ros}, {Nidever}, {Nitschelm}, {Pan}, {Rix},
  {Rybizki}, {Schiavon}, {Schneider}, {Wilson}, \&
  {Zamora}}]{2019ApJ...874..102W}
{Weinberg}, D.~H., {Holtzman}, J.~A., {Hasselquist}, S., {et~al.} 2019, \apj,
  874, 102

\bibitem[{{Weisz} {et~al.}(2015){Weisz}, {Johnson}, {Foreman-Mackey},
  {Dolphin}, {Beerman}, {Williams}, {Dalcanton}, {Rix}, {Hogg}, {Fouesneau},
  {Johnson}, {Bell}, {Boyer}, {Gouliermis}, {Guhathakurta}, {Kalirai}, {Lewis},
  {Seth}, \& {Skillman}}]{2015ApJ...806..198W}
{Weisz}, D.~R., {Johnson}, L.~C., {Foreman-Mackey}, D., {et~al.} 2015, \apj,
  806, 198

\bibitem[{Zeghal {et~al.}(2022)Zeghal, Lanusse, Boucaud, Remy, \&
  Aubourg}]{zeghal2022neuralposteriorestimationdifferentiable}
Zeghal, J., Lanusse, F., Boucaud, A., Remy, B., \& Aubourg, E. 2022, Neural
  Posterior Estimation with Differentiable Simulators

\bibitem[{{Zhou} {et~al.}(2024){Zhou}, {Radev}, {Oliver}, {Obreja}, {Jin}, \&
  {Buck}}]{zhou2024}
{Zhou}, L., {Radev}, S.~T., {Oliver}, W.~H., {et~al.} 2024, arXiv e-prints,
  arXiv:2410.10606

\end{thebibliography}

\begin{appendix}

\section{Code and Data Availability}
\label{sec:appendix_code_and_data}

To facilitate a wider community's usage and contributions, we make use of the reproducibility software
\href{https://github.com/showyourwork/showyourwork}{\showyourwork}
\citep{Luger2021}, which leverages continuous integration to
programmatically download the data from
\href{https://zenodo.org/}{zenodo.org}, create the figures, and
compile the manuscript. Each figure caption contains two links: one
to the dataset stored on zenodo used in the corresponding figure,
and the other to the script used to make the figure (at the commit
corresponding to the current build of the manuscript). The git
repository associated to this study is publicly available at
\url{https://github.com/TobiBu/sbi-chempy}, and the release
v0.4.1 allows anyone to re-build the entire manuscript including rerunning all analysis. The datasets and neural network weights are stored at \url{https://zenodo.org/records/14925307}. The training and validation data can be found on zenodo as well under this link: \url{https://zenodo.org/records/14507221}. 


\begin{figure}[]
     \centering
     \includegraphics[width=\columnwidth]{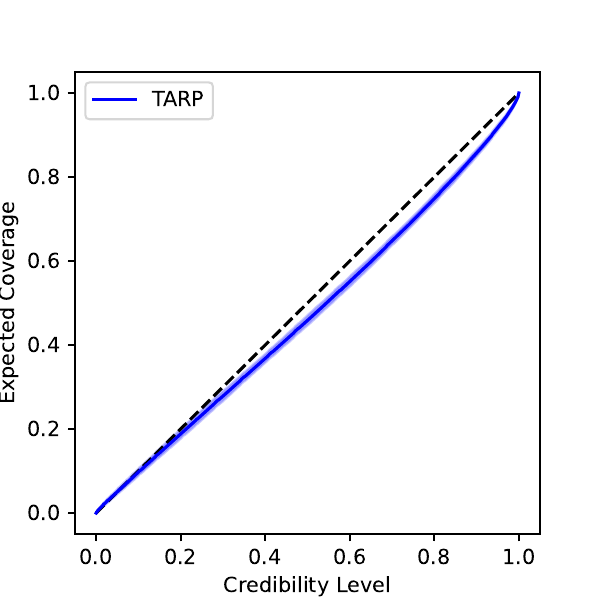}
     \caption{TARP plot showing the expected coverage probability vs. the credibility level $\alpha$. The dashed black 1:1-line shows an ideal calibrated posterior and the blue solid line shows the TARP value for our NPE.}
     \label{fig:tarp}
     \script{plot_sbc.py}
\end{figure}

\begin{figure*}[]
     \centering
     \includegraphics[width=\textwidth, trim={4cm, 0cm, 4cm, 0cm}, clip]{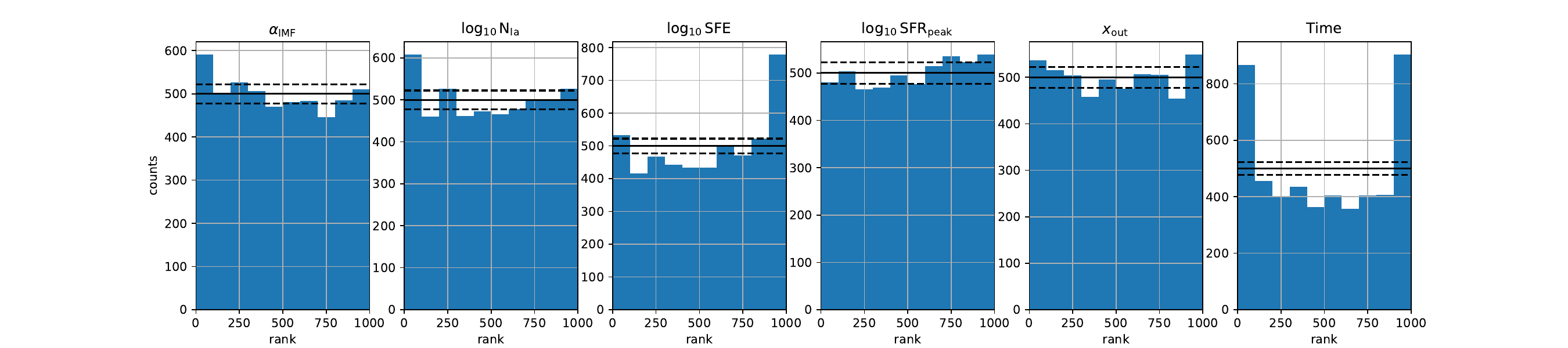}
     \includegraphics[width=\textwidth, trim={6cm, 0cm, 6cm, 0cm}, clip]{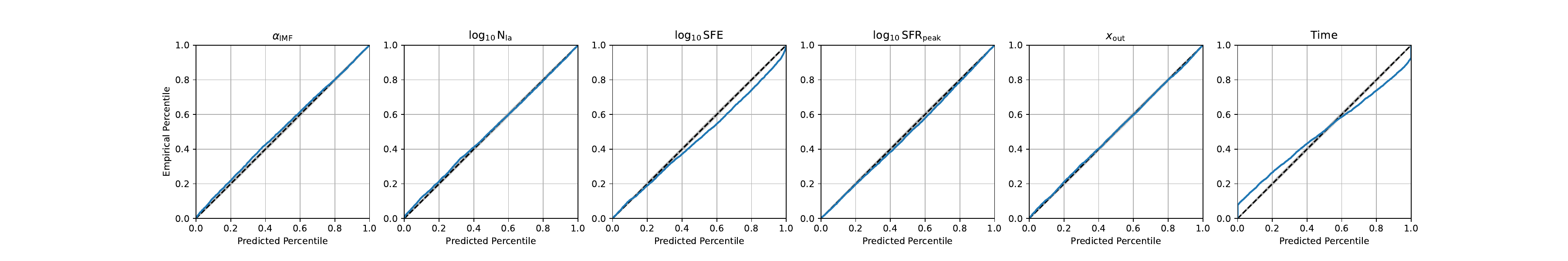}
     \includegraphics[width=\textwidth, trim={6cm, 0cm, 6cm, 0cm}, clip]{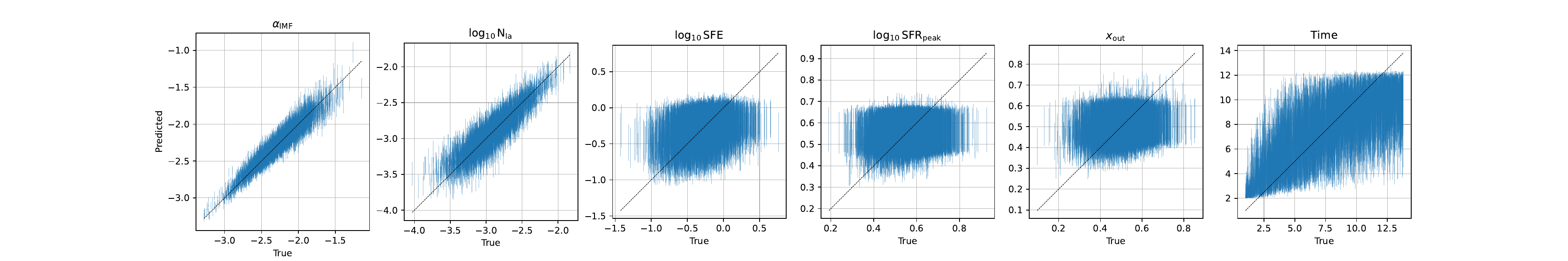}
     \caption{Posterior calibration diagnostics showing from top to bottom the SBC ranks, the TARP plots and the true vs. predicted parameter plots for each of the six parameters.
     Top: SBC ranks of ground truth parameters under the inferred posterior samples for each of the six parameters (red bars). The grey area shows the 99\% confidence interval of a uniform distribution given the number of samples provided. Middle: TARP plot showing the expected coverage probability vs. the credibility level $\alpha$ for each of the six individual parameters in our inference. The dashed black 1:1-line shows an ideal calibrated posterior and the blue solid line shows the TARP value for our NPE. Bottom: True vs. predicted parameter plots showing the average of the posterior samples and as errorbar the standard deviation of the samples vs their ground truth parameter.}
     \label{fig:tarp}
     \script{plot_sbc.py}
\end{figure*}

\section{Neural Posterior Calibration}
\label{sec:sbc}

Since SBI relies in neural networks to approximate posterior densities one important point is to check that neural network hyperparameters are well chosen and that posterior estimates are trustable.

After a density estimator has been trained with simulated data to obtain a posterior, the estimator should be made subject to several diagnostic tests. This needs to be performed before being used for inference given the actual observed data. Posterior Predictive Checks provide one way to "critique" a trained estimator based on its predictive performance. Another important approach to such diagnostics is simulation-based calibration as developed by \citet{Cook2006} and \citet{Talts2018}. 

\paragraph{Simulation-based calibration}
Simulation-based calibration (SBC) provides a (qualitative) view and a quantitive measure to check, whether the variances of the posterior are balanced, i.e. it is neither over-confident nor under-confident. As such, SBC can be viewed as a necessary condition (but not sufficient) for a valid inference algorithm: If SBC checks fail, this tells you that your inference is invalid. If SBC checks pass, this is no guarantee that the posterior estimation is working.

To perform SBC, we sample some $\theta_i^o$ values from the parameter prior of the problem at hand and simulate "observations" from these parameters: 
\begin{equation}
    x_i = \text{simulator}(\theta_i^o)
\end{equation}
Then we perform inference given each observation $x_i$ which produces a separate posterior $p_i(\theta|x_i)$ for each $x_i$. The key step for SBC is to generate a set of posterior samples $\{\theta\}_i$ from each posterior. We call this $\theta_i^s$, referring to $s$ samples from the posterior $p_i(\theta|x_i)$. Next, we rank the corresponding $\theta_i^o$ under this set of samples. A rank is computed by counting how many samples $\theta_i^s$ fall below their corresponding $\theta_i^o$ value \citep[see section 4.1 in][]{Talts2018}. These ranks are then used to perform the SBC check itself.

The core idea behind SBC is two fold: (i) SBC ranks of ground truth parameters under the inferred posterior samples follow a uniform distribution (If the SBC ranks are not uniformly distributed, the posterior is not well calibrated); and (ii) samples from the data averaged posterior (ensemble of randomly chosen posterior samples given multiple distinct observations $x_o$) are distributed according to the prior.

Hence, SBC can tell us whether the SBI method applied to the problem at hand produces posteriors that have well-calibrated uncertainties, and if the posteriors have uncalibrated uncertainties, SBC surfaces what kind of systematic bias is present: negative or positive bias (shift in the mean of the predictions) or over- or under-dispersion (too large or too small variance).

In the top panel of Fig.~\ref{fig:tarp} we show the distribution of ranks (depicted in red) in each dimension. Highlighted with black lines, you see the 99\% confidence interval of a uniform distribution given the number of samples provided. In plain english: for a uniform distribution, we would expect 1 out of 100 (blue) bars to lie outside the grey area. This figure shows that overall our posteriors are decently calibrated. Only for the parameter $\log_{10}\left(\mathrm{SFE}\right)$ and Time we see a slight bmiss-calibration. But most importantly for the parameters of interest here, $\log_{10}\left(N_\mathrm{Ia}\right)$ and $\alpha_\mathrm{IMF}$ we have a well calibrated posterior.

\paragraph{Tests of Accuracy with Random Points (TARP)}

TARP \citep{Lemos2023} is an alternative calibration check for the joint distribution, for which defining a rank is not straightforward. Given a test set $(\theta^*,x^*)$ and a set of reference points $\theta_r$, TARP calculates statistics for posterior calibration by - drawing posterior samples $\theta$ given each observation $x^*$ and calculating the distance $r$ between $\theta^*$ and $\theta_r$ counting for how many of the posterior samples the distance to $\theta_r$ is smaller than $r$ \citep[see e.g. Fig.~2 in][for an illustration]{Lemos2023}.

For each given coverage level $\alpha$, one can then calculate the corresponding average counts and check, whether they correspond to the given $\alpha$. The visualization and interpretation of TARP values is therefore similar to that of SBC. However, in contrast to SBC, TARP provides a necessary and sufficient condition for posterior accuracy, i.e., it can also detect inaccurate posterior estimators. In the middle row of Fig.~\ref{fig:tarp} we show the result for our NPE in blue in comparison to the ideal line shown in black dashed style. This figure clearly shows that our NPE is well calibrated.

Note, however, that this property depends on the choice of reference point distribution: to obtain the full diagnostic power of TARP, one would need to sample reference points from a distribution that depends on $x$. Thus, in general, it is recommended using and interpreting TARP like SBC and complementing coverage checks with posterior predictive checks.

Finally, the bottom row of Fig.~\ref{fig:tarp} shows the predicted vs. true parameter plots where blue dots show the average of the posterior samples and errorbars show the standard deviation of the samples vs their ground truth parameter. We find that for the global parameters we recover the 1:1 relation well while for the other parameters the agreement is tilted.

\section{Correlation analysis of inference results}
\label{sec: correlations}
\begin{figure*}
    \centering
    \includegraphics[width=\textwidth]{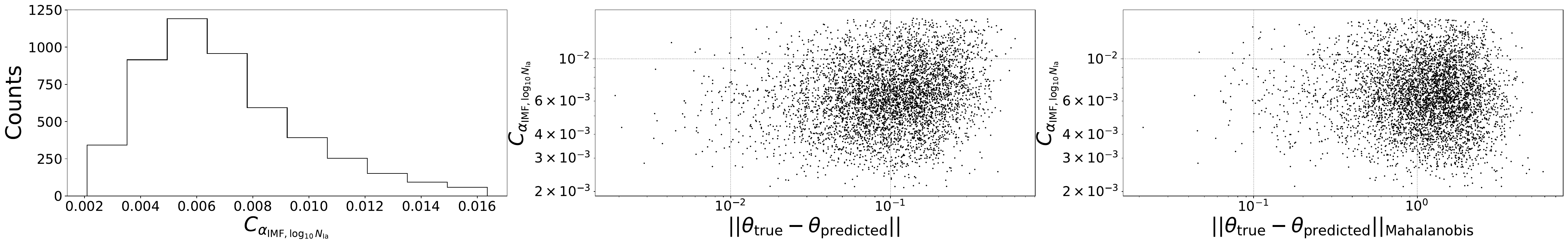}
    \vspace{-.5cm}
    \caption{The left Figure shows the distribution of the correlation between the global parameters $\alpha_{\text{IMF}}$ and $\log_{10} N_{\text{Ia}}$, obtained by the covariance matrix of the sample of each star in the validation set. The central and right Figures show the correlation as a function of the Euclidean and Mahalanobis distance of the true value $\theta_{\text{true}}$ and the sample average $\theta_{\text{predicted}}$. This results shows that the correlation is independent on the accuracy of the inference.}

    \label{fig:correlation}
    \script{plot_sbc.py}
\end{figure*}

In order to characterize the relation between the global parameters $\alpha_{\text{IMF}}$ and $\log_{10} N_{\text{Ia}}$, we have decided to study the correlation obtained from the inference results on the whole validation set. Figure \ref{fig:corner_plot} display a possible positive correlation, and in order to obtain a population level statics of this results, we calculate for each star in the validation set the 2x2 covariance matrix\footnote{We have used only the global parameters components of the sample to calculate the covariance matrix.} of the posterior samples, extracting the off diagonal value. In Figure \ref{fig:correlation}, the histogram of the correlation value is shown, and all the inference results suggest a positive correlation. We kept only the left sided 99 percentile for graphical reasons, but also those removed outliers are in agreement with the conclusion. 
We checked the independence of this results from the accuracy of the inference with the central and right plots of Fig. \ref{fig:correlation}. 

\section{Additional inference results}
\label{sec: additional inference}

\begin{figure*}
    \centering
    \includegraphics[width=\textwidth]{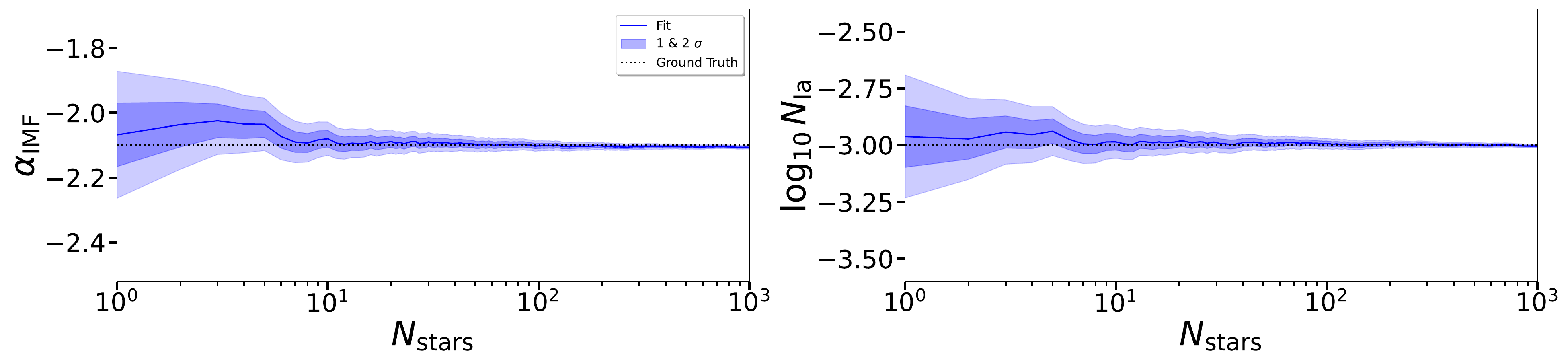}
    \vspace{-.5cm}
    \caption{Same as Fig.~\ref{fig:CHEMPY_TNG_N_star_analysis} but for mock data created with different parameters for $\alpha_{IMF}=-2.1$ and $\log_{10}(N_{Ia})=-3.0$.}
    \label{fig:N_star_analysis_different_prior}
    \script{additional_inference.py}
\end{figure*}

\begin{figure}
    \centering
    \includegraphics[width=\columnwidth]{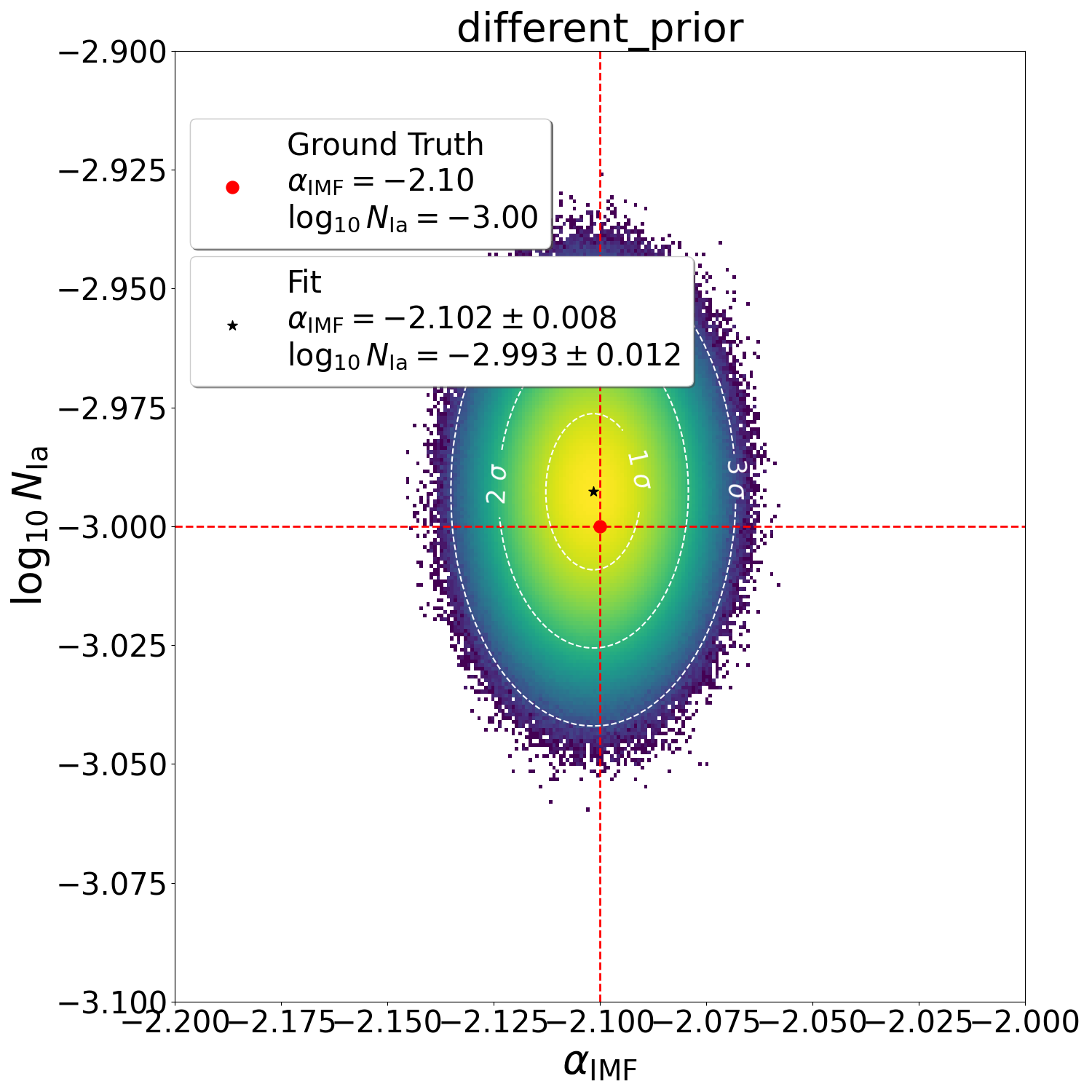}
    \vspace{-.25cm}
    \caption{Same as Fig.~\ref{fig:CHEMPY_TNG_sim_sbi} but for the mock data taken created with a different parameter combination for $\alpha_{IMF}=-2.1$ and $\log_{10}(N_{Ia})=-3.0$.}
    \label{fig:CHEMPY_different_prior_sbi} 
    \script{additional_inference.py}
\end{figure}

\end{appendix}

\end{document}